\documentclass[showpacs,preprintnumbers,amsmath,amssymb,nofootinbib]{revtex4}

\usepackage{setspace}
\usepackage{graphicx}  
\usepackage{dcolumn}   
\usepackage{bm}        
\usepackage{rotating}

\newcommand{\ptjet}{p_{ T}}

\newcommand{\phijet}{\phi}
\newcommand{\rapjet}{y}

\newcommand{\akt}{\hbox{anti-${k_t}$} }
\newcommand{\njet}{N^{\rm jet}}

\begin{document}


\title{Study of Jet Shapes in Inclusive Jet Production in
$p p$ Collisions at $\sqrt{s} = {\rm 7 \ TeV}$ using the ATLAS Detector}
\author{(The ATLAS Collaboration)}
\date{\today}

\begin{abstract}
Jet shapes have been measured in inclusive jet production 
in proton-proton collisions at $\sqrt{s}= 7$~TeV using 3~pb${}^{-1}$
of data recorded by the ATLAS experiment at the LHC.
Jets are reconstructed 
using the \akt algorithm with transverse momentum $30$~GeV $< \ptjet < 600$~GeV 
and rapidity in the region $|\rapjet|<2.8$. The data  are corrected for detector effects   
and compared to several leading-order QCD matrix elements 
plus parton shower Monte Carlo predictions, including different sets of parameters tuned to 
model fragmentation processes and underlying event contributions in the final state.
The measured jets become narrower with increasing jet transverse momentum and the jet shapes present a 
moderate jet rapidity dependence. 
Within QCD, the data test a variety of perturbative and non-perturbative effects.
In particular, the data show sensitivity to the details of the parton shower, fragmentation,  
and underlying event models in the Monte Carlo generators. For an appropriate choice of the 
parameters used in these models, the data are well described.
\end{abstract}


\maketitle


\section{Introduction}

The study of the jet shapes~\cite{jetshape} in proton-proton collisions provides 
information about the details of the parton-to-jet fragmentation process, leading to collimated flows 
of particles in the final state. The internal structure of sufficiently energetic jets is 
mainly dictated by the emission of  multiple gluons from the primary parton, 
calculable in perturbative QCD (pQCD)~\cite{pqcd}. The shape of the jet depends on the type of partons 
(quark or gluon) that give rise to jets in the final state~\cite{bjet},
 and is also sensitive to non-perturbative fragmentation 
effects and underlying event (UE) contributions from the interaction between
proton remnants. A proper modeling of the soft contributions is crucial 
for the understanding of  jet production in hadron-hadron collisions 
and for the comparison of the jet cross section measurements with  pQCD  
theoretical predictions~\cite{jet_pro,jet_pro_atlas}. In addition, jet shape related 
observables have been recently proposed~\cite{boosted} to search for new physics in event 
topologies with highly boosted particles in the final state decaying into multiple jets of particles.   

Jet shape measurements have previously been performed in 
$p\bar{p}$~\cite{ppbar}, $e^{\pm}p$~\cite{ep}, and $e^+ e^-$~\cite{ee} collisions. 
In this paper, measurements of differential and integrated jet shapes 
in proton-proton  collisions at  
$\sqrt{s}=7$~TeV are presented for the first time. The study uses data 
collected by the  ATLAS experiment corresponding to $3$~pb${}^{-1}$ of total integrated luminosity.
The measurements are corrected for detector effects and compared to several Monte Carlo (MC)
predictions based on pQCD leading-order (LO) matrix elements plus parton showers, and
including different phenomenological models to describe fragmentation processes and UE contributions.


The paper is organised as follows. The detector is described in the next section. Section~3 
discusses the simulations used in the measurements, while Section~4 and Section~5 provide details 
on jet reconstruction and event selection, respectively. Jet shape observables are defined in Section~6. 
The procedure used to correct the measurements for detector effects is explained in Section~7, 
and the study of systematic uncertainties is discussed in Section~8. The jet shape measurements are presented
in Section~9. Finally, Section~10 is devoted to summary and conclusions.


\section{Experimental setup}

The ATLAS detector~\cite{atlas} 
covers nearly the entire solid angle around
the collision point with layers of tracking detectors, calorimeters, and muon
chambers. For the measurements presented in this paper, 
the tracking system and calorimeters are of particular importance.

The ATLAS inner detector has full coverage  in $\phi$~\cite{coord}
and covers the pseudorapidity range $|\eta|<2.5$. It consists of a silicon pixel detector, a silicon microstrip detector and a transition radiation tracker, all  immersed in a 2 Tesla magnetic field.
High granularity liquid-argon (LAr) electromagnetic sampling calorimeters cover the pseudorapidity
range $|\eta|<$~3.2. The hadronic calorimetry in the range $|\eta|<$~1.7 is provided by a scintillator-tile calorimeter, which
is separated into a large barrel and two smaller extended barrel cylinders, one on either side of
the central barrel. In the end-caps ($|\eta|>$~1.5), LAr hadronic
calorimeters match the outer $|\eta|$ limits of the end-cap electromagnetic calorimeters. The LAr
forward calorimeters provide both electromagnetic and hadronic energy measurements, and they extend
the coverage to $|\eta| < 4.9$.

The trigger system uses three consecutive trigger levels to select   
events. The Level-1 (L1) trigger is based on custom-built hardware to process the incoming data with 
a fixed latency of 2.5~$\mu s$. This is the only trigger level used in this analysis.
The events studied here are selected 
either by the system of minimum-bias trigger
scintillators (MBTS) or by the calorimeter trigger. 
The MBTS detector~\cite{mbts} consists of 32 scintillator counters
of thickness 2~cm organized in two disks. The disks are installed on the 
inner face of the end-cap calorimeter cryostats at $z = \pm 356$~cm, such that the disk surface is perpendicular to 
the beam direction. This leads to a coverage of $2.09 < |\eta| < 3.84$.
The jet trigger is based on the selection of jets according to their transverse energy, $E_T$. 
The L1 jet reconstruction uses the so called jet elements, which are made of electromagnetic and hadronic 
cells grouped together with a granularity of $\Delta\phi \times  \Delta\eta = 0.2 \times 0.2$ for $|\eta|<3.2$. The 
jet finding is based on a sliding window algorithm with steps of one jet element, 
and the jet $E_T$ is computed in a window of configurable size around the jet.


\section{Monte Carlo simulation}

Monte Carlo simulated  samples are used to determine and correct for detector effects, and to 
estimate part of the systematic uncertainties on the measured jet shapes. 
Samples of inclusive jet 
events in proton-proton 
collisions at $\sqrt{s} = 7$~TeV are produced 
using both   PYTHIA~6.4.21~\cite{PYTHIA} and   HERWIG++~2.4.2~\cite{HERWIG++} event generators.
These MC programs implement LO pQCD matrix elements for $2 \rightarrow 2$ processes plus parton shower in the leading logarithmic approximation,  and the string~\cite{string} and cluster~\cite{cluster} models for fragmentation into hadrons, respectively. 
In the case of   PYTHIA, different
MC samples with slightly different parton
shower and UE modeling in the final state are
considered. The samples are 
generated using three  tuned sets of parameters denoted as 
  ATLAS-MC09~\cite{MC09},   DW~\cite{DW}, and   Perugia2010~\cite{Perugia2010}.
In addition, a special PYTHIA-Perugia2010 sample without UE contributions is generated.  
Finally, 
inclusive jet samples are also produced using the ALPGEN 2.13~\cite{ALPGEN} event generator 
interfaced with HERWIG 6.5~\cite{herwig} and JIMMY 3.41~\cite{Jimmy} to model the UE contributions.  
HERWIG++ and PYTHIA-MC09 samples are generated with MRST2007LO${}^{*}$~\cite{mrst2007lo} parton density functions (PDFs) inside the 
proton, PYTHIA-Perugia2010 and PYTHIA-DW with CTEQ5L~\cite{cteq} PDFs, and 
ALPGEN with CTEQ61L~\cite{cteq6} PDFs. 

The MC generated samples are passed through a full simulation~\cite{atlas_sim} 
of the ATLAS detector 
and trigger, based on GEANT4~\cite{geant}.
The Quark Gluon String Precompound (QGSP) model~\cite{qgs} is used for the fragmentation of the nucleus, and 
the Bertini cascade (BERT) model~\cite{bertini} 
for the description 
of the interactions of the hadrons in the medium of the nucleus. 
Test-beam measurements for single pions have shown that these simulation settings best describe 
the response and resolution in the barrel~\cite{sim_barrel} and 
end-cap~\cite{sim_endcap} calorimeters.  
The simulated events are then reconstructed and
analyzed with the same analysis chain as for the data, and the same trigger and event 
selection criteria. 

\section{Jet reconstruction}

Jets are defined using the \akt jet algorithm~\cite{akt} with distance 
parameter (in $y - \phi$ space)  $R=0.6$, and the 
energy depositions in calorimeter clusters as input in both data and MC  events. 
Topological clusters~\cite{jet_pro_atlas} are built around seed calorimeter cells with $|E_{\rm cell}|>4\sigma$,
 where $\sigma$ is defined as the RMS of the cell energy noise distribution, to which all directly neighboring cells are added. Further neighbors of neighbors are iteratively added for all cells with signals above a secondary threshold $|E_{\rm cell}|>2\sigma$, and the clusters are set massless. 
In addition, in the simulated events jets are also defined at the particle level~\cite{particle} using as 
input all the final state particles from the MC generation.

The \akt algorithm constructs, for each input object (either energy cluster or particle) $i$, 
the quantities $d_{ij}$ and $d_{iB}$ as follows:
\begin{eqnarray}
d_{ij} &=& \min(k_{t i}^{-2},k_{t j}^{-2})\frac{(\Delta R)^2_{ij}}{R^2}, \\
d_{iB} &=& k_{t i}^{-2},
\label{eq:dist}
\end{eqnarray} 
where
\begin{equation}
(\Delta R)^2_{ij} = (y_i - y_j)^2 + (\phi_i - \phi_j)^2, 
\end{equation}
$k_{t i}$ is the transverse momentum of object $i$ with respect to the beam direction, $\phi_i$ its 
azimuthal angle,
and $y_i$ its rapidity.
A list containing all the $d_{ij}$ and $d_{iB}$ values is compiled. If the smallest entry is a $d_{ij}$, 
objects $i$ and $j$ are 
combined (their four-vectors are added) and the list is updated. If the smallest entry is a $d_{iB}$, 
this object is considered a complete 
``jet'' and is removed from the list. As defined above, $d_{ij}$ is a distance measure between
two objects, and $d_{iB}$ is a similar distance between the object and the beam. Thus the variable $R$ is a resolution
parameter which sets the relative distance at which jets are resolved from each other as compared to the beam. 
The \akt algorithm is theoretically well-motivated~\cite{akt} and produces geometrically well-defined (``cone-like'') jets. 

According to MC simulation, the measured jet angular variables, $\rapjet$ and $\phijet$, are reconstructed with a resolution of better than 0.05 units, which improves as the jet transverse momentum, $\ptjet$, increases. The measured jet $\ptjet$ 
is corrected to the particle level scale~\cite{jet_pro_atlas} using an average correction, computed as a function of jet transverse momentum and pseudorapidity, and extracted from MC simulation.

\section{Event selection}

The data were collected during the first LHC run at $\sqrt{s} = 7$~TeV 
with the ATLAS  tracking detectors, calorimeters and magnets 
operating at nominal conditions.
Events are selected online using different L1 trigger 
configurations in such a way that, in  the kinematic 
range for the jets considered in this study (see below), the  trigger 
selection is fully efficient and does not introduce any significant bias in the measured jet shapes. 
Table~1 presents the trigger configurations employed in each $\ptjet$ region and the 
corresponding integrated luminosity. The unprescaled trigger thresholds were increased with time 
to keep pace with the LHC instantaneous luminosity evolution.
For jet $\ptjet$ smaller than 60~GeV, the data are selected using
the signals from the MBTS detectors on either side of the 
interaction point.
Only events in which the MBTS recorded one or more counters 
above threshold on at least one side are retained. 
For larger $\ptjet$, the events are selected   
using either MBTS or L1 calorimeter based triggers (see Section~2) with a minimum transverse 
energy threshold at the electromagnetic scale~\cite{emscale} 
that varies between 5~GeV (L1\_5) and 55~GeV (L1\_55),  depending on when the data were collected and the  
$\ptjet$ range considered (see Table~1).

The events are required to  
have one and only one reconstructed primary vertex with 
a $z$-position within 10~cm of the origin of the coordinate system, which
suppresses pile-up contributions from multiple proton-proton interactions in the same bunch crossing, beam-related backgrounds and cosmic rays.
In this analysis, events are required to have at least one jet with corrected 
transverse momentum $\ptjet > 30$~GeV and rapidity $|\rapjet| < 2.8$. This corresponds approximately to 
the kinematic region, in the absolute four momentum transfer squared $Q^2$ - Bjorken-$x$ plane, of  $10^3$~GeV${}^2  < Q^2 < 4 \times 10^5$~GeV${}^2$ and $6 \times 10^{-4} < x < 2 \times 10^{-2}$. Additional quality criteria are  applied to ensure that jets are not produced by 
noisy calorimeter cells, and to avoid problematic detector regions.

\begin{table}
\begin{small}
\begin{center}
\begin{tabular}{|c|l|c|} \hline
\multicolumn{3}{|c|}{{\small{Trigger Information}}} \\ \hline\hline\hline
$\ptjet$ (GeV) & trigger configurations & integrated luminosity (nb${}^{-1})$ \\ \hline\hline
30 - 60 &   MBTS & 0.7\\
60 - 80 &   L1\_5/MBTS & 17 \\
80 - 110 &  L1\_10/L1\_5/MBTS & 96\\
110 - 160 & L1\_15/L1\_10/L1\_5/MBTS & 545\\
160 - 210 & L1\_30/L1\_15/L1\_10/L1\_5/MBTS & 1878\\ 
210 - 600 & L1\_55/L1\_30/L1\_15/L1\_10/L1\_5/MBTS & 2993  \\ \hline\hline
\end{tabular}
\label{tab:trigger}
\caption{\small
For the various jet $\ptjet$ ranges, the trigger configurations used to collect the data
and the corresponding total integrated luminosity. MBTS denotes the use of the minimum-bias trigger 
scintillators, while L1\_5, L1\_10, L1\_15, L1\_30, and L1\_55 correspond to L1 calorimeter triggers 
with 5, 10, 15, 30, and 55 GeV thresholds, respectively.   
}
\end{center}
\end{small}
\end{table}


\section{Jet shape definition}

The internal structure of the jet is studied in terms of the differential and integrated 
jet shapes, as  reconstructed using the uncorrected energy clusters in the calorimeter  associated with the jet. 
The differential jet shape $\rho(r)$ as a function of
the distance 
 $r=\sqrt{\Delta y^2 + \Delta \phi^2}$ to the jet axis is
defined as the average fraction of the 
jet $\ptjet$ that lies inside an annulus of inner radius
${r - \Delta r/2}$ and outer radius ${r + \Delta r/2}$ around the jet axis:

\begin{equation}
\rho (r) = \frac{1}{\Delta r}\frac{1}{\njet}\sum_{\rm jets}\frac{p_{T}(r-\Delta r/2,r+\Delta r/2)}{p_{T}
(0,R)},\ \Delta r/2 \le r \le R-\Delta r/2, 
\end{equation}

\noindent
where $  p_{T}(r_1,r_2)$ denotes the summed $p_{T}$ of the clusters in the annulus between 
radius $r_1$ and $r_2$, $\njet$ is the number of jets, and $R=0.6$ and $\Delta r = 0.1$ are used. 
 The points from the differential jet shape 
at different $r$ values are correlated since, by definition, $\sum_0^R \rho (r) \ \Delta r = 1$. 
Alternatively, the integrated jet shape $\Psi(r)$ is defined as the 
average fraction of the jet $\ptjet$ that lies inside a cone of radius $r$ 
concentric with the jet cone:

\begin{equation}
\Psi (r) = \frac{1}{\njet}\sum_{\rm jets}\frac{p_{T}(0,r)}{p_{T}(0,R)},\ 0 \le r \le R, 
\end{equation}

\noindent
where, by definition, $\Psi (r = R) = 1$, and the points at different $r$ values are correlated. 
The same definitions apply to simulated calorimeter clusters and 
final-state particles in  the MC generated events to define
differential and integrated jet shapes at the calorimeter and particle levels, respectively. 
The jet shape measurements are performed in different regions of jet $\ptjet$ and $|\rapjet|$, and a minimum 
of 100 jets in data are required in each region to limit the statistical 
fluctuations on the measured values.


\section{Correction for detector effects}

The measured differential and integrated jet shapes, as determined by using calorimeter topological clusters, 
are corrected for detector effects back to the particle level. This is done using MC simulated events and a bin-by-bin 
correction procedure that also accounts for the efficiency of the selection 
criteria and of the jet reconstruction in the calorimeter. 
  PYTHIA-Perugia2010 provides a 
reasonable description of the measured jet shapes in all regions of jet $\ptjet$ and $|\rapjet|$,  
and is therefore used to compute the correction factors.
Here, the method is described in detail for the differential case.
A similar procedure is employed to correct independently the integrated measurements.
The correction factors $U(r,\ptjet,|\rapjet|)$ are computed 
separately in each jet $\ptjet$ and $|\rapjet|$ region. They are defined as the ratio between the 
jet shapes at the particle level  $\rho(r)^{par}_{mc}$, 
obtained using particle-level jets in the kinematic range under consideration, 
and the reconstructed jet shapes at the 
calorimeter level  $\rho(r)^{cal}_{mc}$, after the selection criteria are applied and using calorimeter-level jets
in the given $\ptjet$ and $|\rapjet|$ range.
The correction factors $U(r,\ptjet,|\rapjet|) = \rho(r)^{par}_{mc}/\rho(r)^{cal}_{mc}$ present a moderate $\ptjet$ and $|\rapjet|$
dependence and vary between 0.95 and 1.1 as $r$ increases. For the integrated jet shapes, the correction factors differ from unity by less than $5 \%$. The corrected jet shape measurements in each $\ptjet$ and $|\rapjet|$ region 
are computed by multiplying bin-by-bin the measured uncorrected jet shapes in data 
by the corresponding correction factors.


\section{Systematic uncertainties}

A detailed study of systematic uncertainties on the measured differential and integrated 
jet shapes has been performed.
The impact on the differential measurements is described here in detail.

\begin{itemize}

\item The absolute energy scale of the individual clusters belonging to the jet
is varied in the data according to studies using
isolated tracks~\cite{jet_pro_atlas}, which parametrize the uncertainty on
the calorimeter cluster energy as a function of $\ptjet$ and $\eta$ 
of the cluster. This introduces a systematic
uncertainty on the measured differential jet shapes that varies
between 3$\%$ to 15$\%$ as $r$ increases and constitutes the dominant 
systematic uncertainty in this analysis.

\item The systematic uncertainty on the measured jet shapes  
arising from the details of the model used to simulate calorimeter showers in the MC  events is
studied.  A different simulated sample is considered, where the
FRITIOF~\cite{ftfp} plus BERT showering model is employed instead of the QGSP plus BERT model.
FRITOF+BERT provides the second best description of the test-beam results~\cite{sim_barrel} after QGSP+BERT.
This introduces an uncertainty on the measured differential jet shapes that varies
between $1\%$ to 4$\%$, and is approximately independent of  $\ptjet$
and $|\rapjet|$.

\item The measured jet $\ptjet$ is varied by 2$\%$ to 8$\%$,  
depending on $\ptjet$ and $|\rapjet|$, to account for the remaining 
uncertainty on the absolute jet energy scale~\cite{jet_pro_atlas},  
after removing contributions already accounted for and related to the 
energy of the single clusters and  the calorimeter shower modeling, as discussed above.
This introduces an uncertainty of about 3$\%$ to 5$\%$ 
in the measured differential jet shapes. 

\item The 14$\%$ uncertainty 
on the jet energy resolution~\cite{jet_pro_atlas} translates into a smaller than 2$\%$
effect on the measured differential jet shapes.

\item The correction factors are recomputed using HERWIG++,
which implements different parton shower, fragmentation
and UE models than PYTHIA, and compared to   PYTHIA-Perugia2010.
In addition, the correction factors 
are also computed using ALPGEN and PYTHIA-DW for $\ptjet < 110$~GeV, where these MC 
samples provide a reasonable description of the uncorrected shapes in the data. 
The results from HERWIG++ encompass the variations
obtained using all the above generators and are conservatively
adopted in all $\ptjet$ and $|\rapjet|$ ranges to compute systematic
uncertainties on the differential jet shapes. These
uncertainties increase between 2$\%$ and 10$\%$ with increasing r.

\item An additional 1$\%$ uncertainty on the differential measurements
is included to account for deviations from unity (non-closure) in the bin-by-bin 
correction procedure when applied to 
a statistically independent MC sample.

\item No significant dependence on instantaneous luminosity
is observed in the measured jet shapes, indicating that residual
pile-up contributions are negligible after selecting events with 
only one reconstructed primary vertex. 

\item It was verified that the presence of small 
dead calorimeter regions in the data does not affect 
the measured jet shapes.  

\end{itemize}

\noindent
The different systematic uncertainties are added in
quadrature to the statistical uncertainty to obtain the final result.
The total uncertainty for differential jet shapes decreases with
increasing $\ptjet$ and varies typically between 3$\%$ and 10$\%$ 
(10$\%$ and 20$\%$) at $r = 0.05$ ($r = 0.55$). The total uncertainty is dominated by
the systematic uncertainty, except at very large $\ptjet$ where the
measurements are still statistically limited.  In the case of the
integrated measurements, the total systematic uncertainty varies
between 10$\%$ and $2\%$ ($4\%$ and $1\%$) at $r=0.1$ ($r=0.3$) as
$\ptjet$ increases, and vanishes as $r$ approaches
the edge of the jet cone. 

Finally, the jet shape analysis is also performed using either  
tracks from the inner detector inside the jet cone, as reconstructed 
using topological clusters; or calorimeter towers of fixed size $0.1 \times 0.1$ ($y - \phi$ space) instead of 
topological clusters as input to the jet reconstruction algorithm. 
For the former, the measurements are limited to jets with $|\rapjet|<1.9$, as dictated by the
tracking coverage and the chosen size of the jet. 
After the data are corrected back to particle level, the results
from these alternative analyses are consistent with the nominal
results, with maximum deviations in the differential measurements of 
about  2$\%$ (5$\%$) at r=0.05 (r=0.55), well within the quoted 
systematic uncertainties.

\section{Results}

The measurements presented in this article refer to differential 
and integrated jet shapes, $\rho (r)$ 
and $\Psi (r)$,  corrected at the particle level and obtained for \akt jets with distance parameter 
$R=0.6$  in the region  $|\rapjet| < 2.8$ and 
$30 \  {\rm{ GeV}} < \ptjet < 600 \ \rm GeV$. The measurements are presented in separate bins of 
 $\ptjet$ and $|\rapjet|$. Tabulated values
of the results are available in the Appendix and in Ref.~\cite{tab}.
 
Figures~\ref{fig1} to ~\ref{fig2} show the measured differential
jet shapes as a function of $r$ in different $\ptjet$ ranges. 
The dominant peak at small $r$ indicates that the 
majority of the jet momentum is concentrated close to the jet axis.  
At low  $\ptjet$,  more than 80$\%$ of the transverse
momentum is contained within a cone of radius $r=0.3$ around the jet direction.
This  fraction increases up to 95$\%$ at very high $\ptjet$, showing 
that jets become narrower as $\ptjet$ increases.  
This is also observed in Fig.~\ref{fig3}, where the measured $1-\Psi(0.3)$, the 
fraction of the jet transverse momentum outside a fixed radius $r=0.3$, decreases as a function of $\ptjet$.  


 The data are
compared to predictions from   HERWIG++, ALPGEN,   PYTHIA-Perugia2010, and
  PYTHIA-MC09 in Fig.~\ref{fig1} to Fig.~\ref{fig3}(a); and to predictions from 
PYTHIA-DW and PYTHIA-Perugia2010 with and without UE contributions in Fig.~\ref{fig3}(b).
The jet shapes predicted by    
  PYTHIA-Perugia2010 provide a reasonable description of the data,
while   HERWIG++ predicts broader jets than the data
at low and very high $\ptjet$.
The PYTHIA-DW predictions are in between
  PYTHIA-Perugia2010 and   HERWIG++ at low $\ptjet$ and produce jets which are 
slightly narrower  at high $\ptjet$. 
ALPGEN is similar to PYTHIA-Perugia2010 at low $\ptjet$, but produces  
jets significantly narrower than the data  
at high $\ptjet$.  PYTHIA-MC09 tends to produce narrower jets  
than the data in the whole kinematic range under study. 
The latter may be  attributed to an
inadequate modeling of the soft gluon radiation and UE
contributions in   PYTHIA-MC09 samples, in agreement with previous
observations of the particle flow activity in the final
state~\cite{mbts}. Finally, Fig.~\ref{fig3}(b) shows that 
PYTHIA-Perugia2010 without UE contributions predicts jets much narrower than the 
data at low $\ptjet$. This  confirms the sensitivity of jet shape observables in the region 
$\ptjet < 160$~GeV to a proper description of the  UE activity in the final state.  
 
The dependence on $|\rapjet|$ is shown in 
Fig.~\ref{fig4}, where the measured jet shapes are presented 
separately in five different jet rapidity regions and 
different $\ptjet$ bins, for jets with $\ptjet < 400$~GeV. 
At high $\ptjet$, the measured  $1-\Psi(0.3)$ shape presents a mild $|\rapjet|$ dependence, indicating that the 
jets become slightly narrower in the forward regions. This tendency is observed also in the various 
MC samples. Similarly, Figs.~\ref{fig5} and \ref{fig6} present the measured
$1-\Psi(0.3)$ as a function of $\ptjet$ in the different  $|\rapjet|$ regions compared 
to PYTHIA-Perugia2010 predictions.  
The result of $\chi^2$ tests to the data in Fig.~\ref{fig6} with respect to the predictions from the 
different  MC generators are reported in Table~7, for each of the five 
rapidity regions. Here 
the different sources of systematic uncertainty are considered independent and fully 
correlated across $\ptjet$ bins (see Appendix). As already discussed,   PYTHIA-Perugia2010 provides 
the best overall description of the data, while PYTHIA-Perugia2010 without UE contributions and ALPGEN 
show the largest discrepancies.    

 Finally, and only for illustration, the typical shapes of   
quark- and gluon-initiated jets, as determined using events generated with PYTHIA-Perugia2010, are also shown in Figs.~\ref{fig5} and ~\ref{fig6}.  
For this purpose, MC events are selected with at least two particle-level jets with $\ptjet > 30$~GeV and $|\rapjet|<2.8$ 
in the final state. 
The two leading jets in this dijet sample are classified as quark-initiated or gluon-initiated jets 
by matching (in $y - \phi$ space) their direction with one of the outgoing partons from the QCD 
$2 \rightarrow 2$ hard process. At low $\ptjet$ the measured jet shapes are 
similar to those from 
gluon-initiated jets, as expected from the dominance of hard processes with gluons in the final state.
At high $\ptjet$,  where the impact of the UE contributions becomes smaller (see Fig.~\ref{fig3}(b)),  the 
observed trend  with $\ptjet$ in the data is mainly attributed to a changing quark- and gluon-jet mixture 
in the final state, convoluted with  perturbative QCD effects related to the running of the strong coupling.
%


\section{Summary and conclusions}

In summary, jet shapes have been measured in inclusive jet production 
in proton-proton collisions at $\sqrt{s}= 7$~TeV using 3~pb${}^{-1}$
of data recorded by the ATLAS experiment at the LHC.
Jets are reconstructed
using the \akt algorithm 
with distance parameter $R=0.6$ in the kinematic region 
30~GeV $< \ptjet < 600$~GeV and $|\rapjet|<2.8$. The data 
are corrected for detector effects and compared to different 
leading-order matrix elements plus parton shower MC
predictions.
The measured jets become narrower as the jet transverse
momentum and rapidity increase, although with a 
rather mild rapidity dependence. The data are reasonably well
described by PYTHIA-Perugia2010. HERWIG++ predicts jets slightly broader than the data, whereas  
ALPGEN interfaced with HERWIG and JIMMY, PYTHIA-DW, and PYTHIA-MC09 all predict jets narrower than the data. 
Within QCD, the data show sensitivity to a  variety of perturbative and non-perturbative effects.
The results reported in this paper indicate the 
potential of jet shape measurements at the LHC  
to constrain  the current phenomenological models for soft gluon radiation, UE 
activity, and non-perturbative fragmentation processes in the final state.




\section{Acknowledgements}

We wish to thank CERN for the efficient commissioning and operation of the LHC during this initial high-energy data-taking period as well as the support staff from our institutions without whom ATLAS could not be operated efficiently.

We acknowledge the support of ANPCyT, Argentina; YerPhI, Armenia; ARC, Australia; BMWF, Austria; ANAS, Azerbaijan; SSTC, Belarus; CNPq and FAPESP, Brazil; NSERC, NRC and CFI, Canada; CERN; CONICYT, Chile; CAS, MOST and NSFC, China; COLCIENCIAS, Colombia; MSMT CR, MPO CR and VSC CR, Czech Republic; DNRF, DNSRC and Lundbeck Foundation, Denmark; ARTEMIS, European Union; IN2P3-CNRS, CEA-DSM/IRFU, France; GNAS, Georgia; BMBF, DFG, HGF, MPG and AvH Foundation, Germany; GSRT, Greece; ISF, MINERVA, GIF, DIP and Benoziyo Center, Israel; INFN, Italy; MEXT and JSPS, Japan; CNRST, Morocco; FOM and NWO, Netherlands; RCN, Norway;  MNiSW, Poland; GRICES and FCT, Portugal;  MERYS (MECTS), Romania;  MES of Russia and ROSATOM, Russian Federation; JINR; MSTD, Serbia; MSSR, Slovakia; ARRS and MVZT, Slovenia; DST/NRF, South Africa; MICINN, Spain; SRC and Wallenberg Foundation, Sweden; SER,  SNSF and Cantons of Bern and Geneva, Switzerland;  NSC, Taiwan; TAEK, Turkey; STFC, the Royal Society and Leverhulme Trust, United Kingdom; DOE and NSF, United States of America.  

The crucial computing support from all WLCG partners is acknowledged gratefully, in particular from CERN and the ATLAS Tier-1 facilities at TRIUMF (Canada), NDGF (Denmark, Norway, Sweden), CC-IN2P3 (France), KIT/GridKA (Germany), INFN-CNAF (Italy), NL-T1 (Netherlands), PIC (Spain), ASGC (Taiwan), RAL (UK) and BNL (USA) and in the Tier-2 facilities worldwide.



\clearpage


\begin{figure}[tbh]
\begin{center}
\mbox{
\includegraphics[width=0.5\textwidth,height=0.52\textwidth]{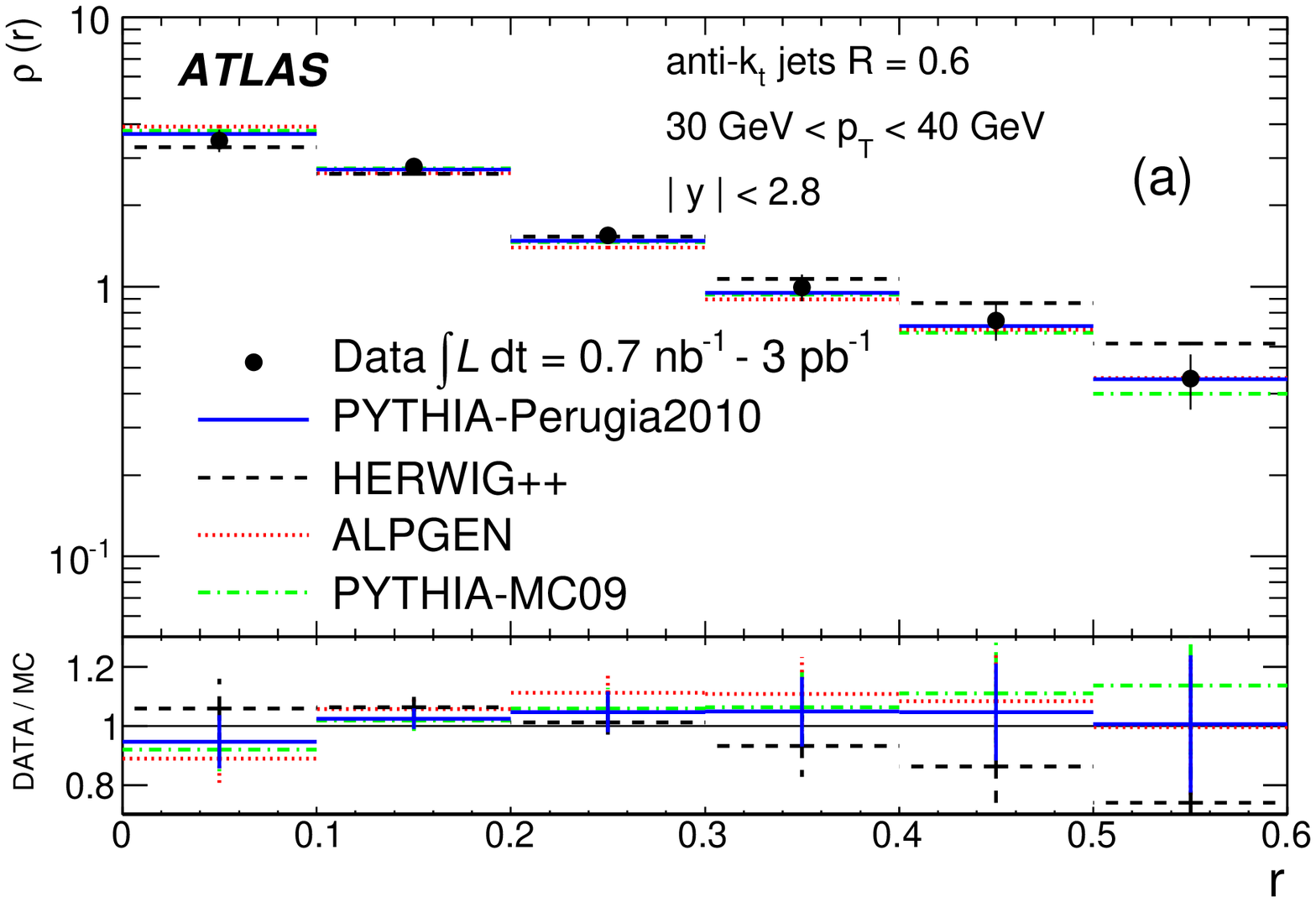} 
\includegraphics[width=0.5\textwidth,height=0.52\textwidth]{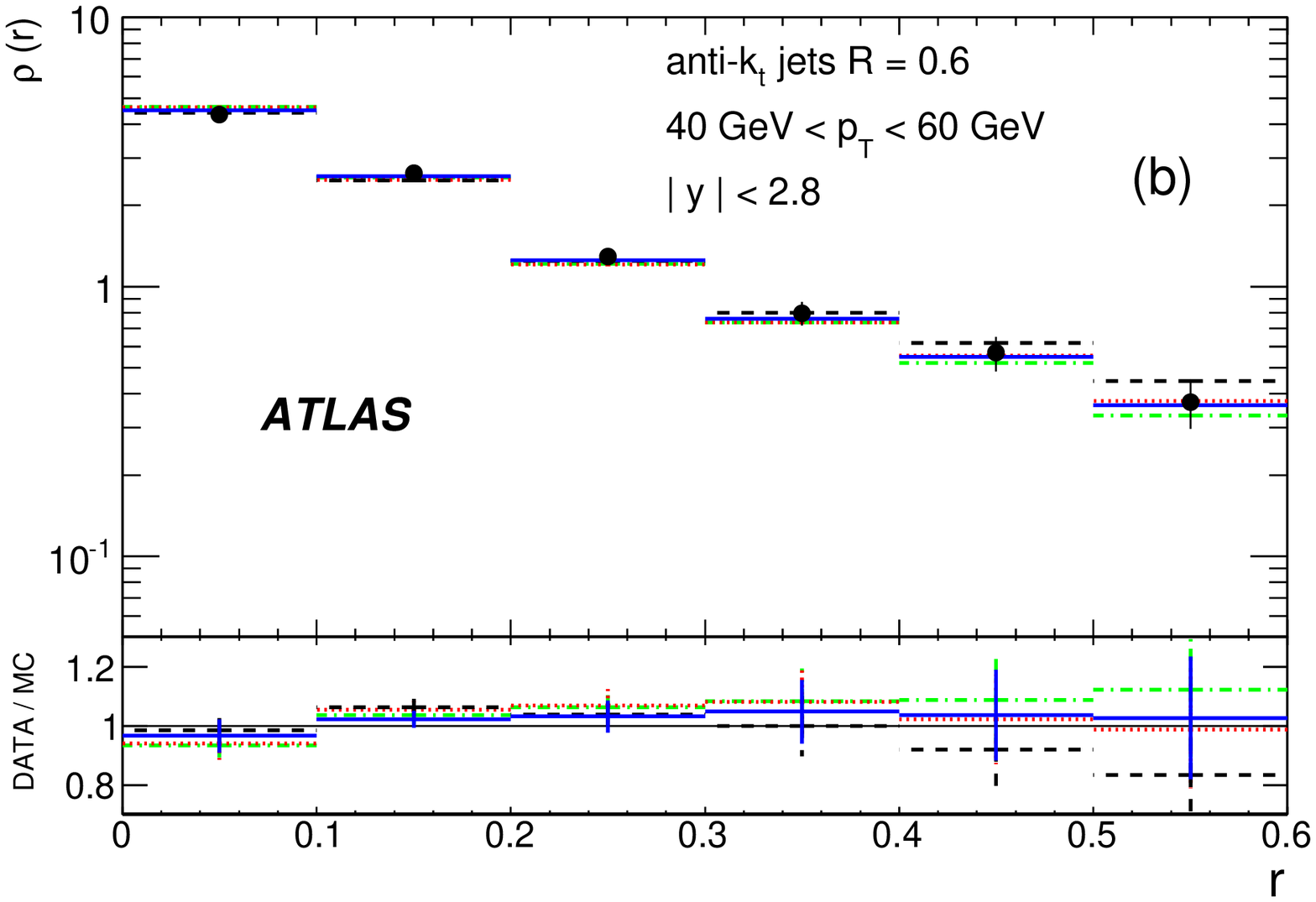}
}\vspace{-0.2cm}
\mbox{
\includegraphics[width=0.5\textwidth,height=0.52\textwidth]{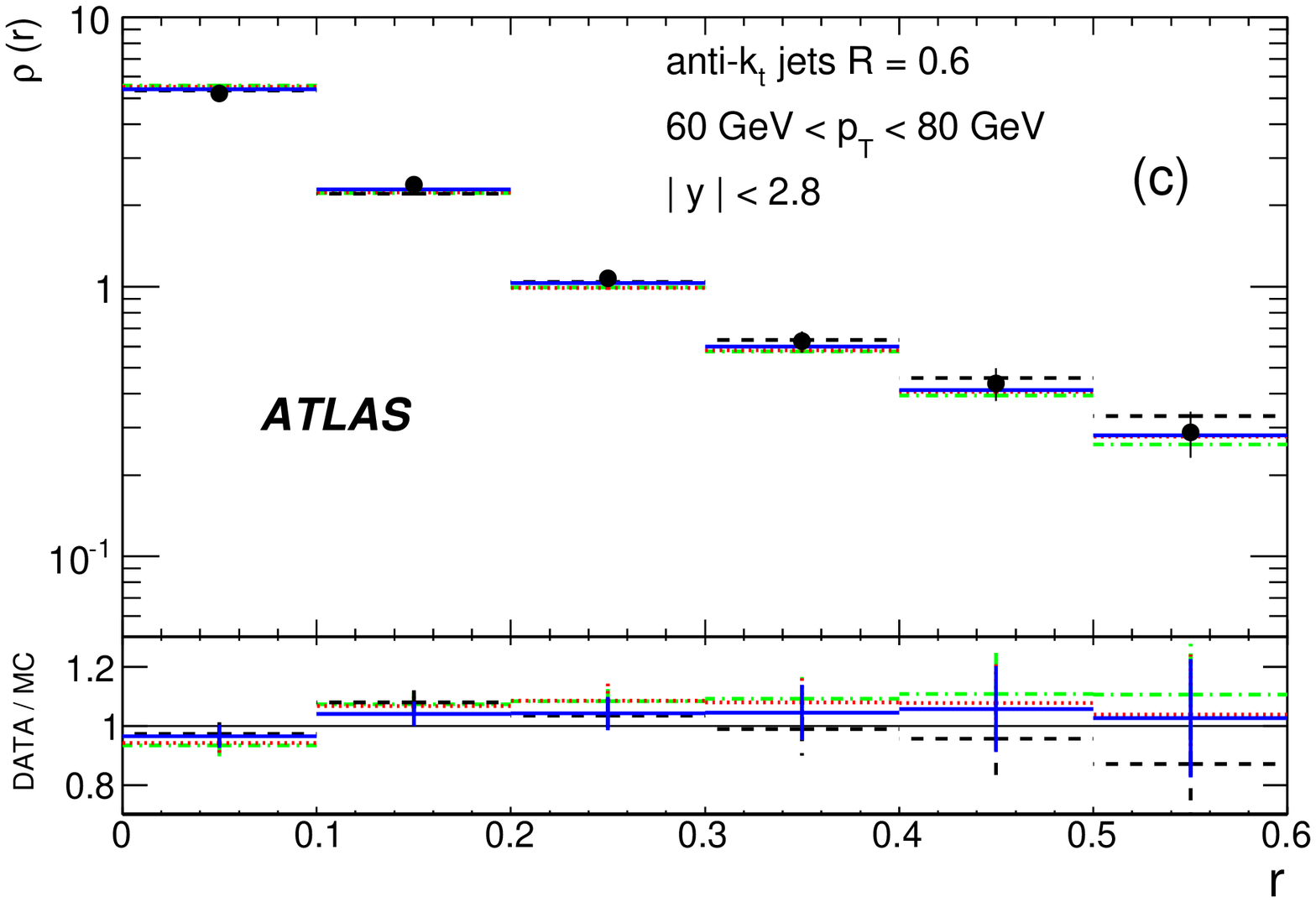}
\includegraphics[width=0.5\textwidth,height=0.52\textwidth]{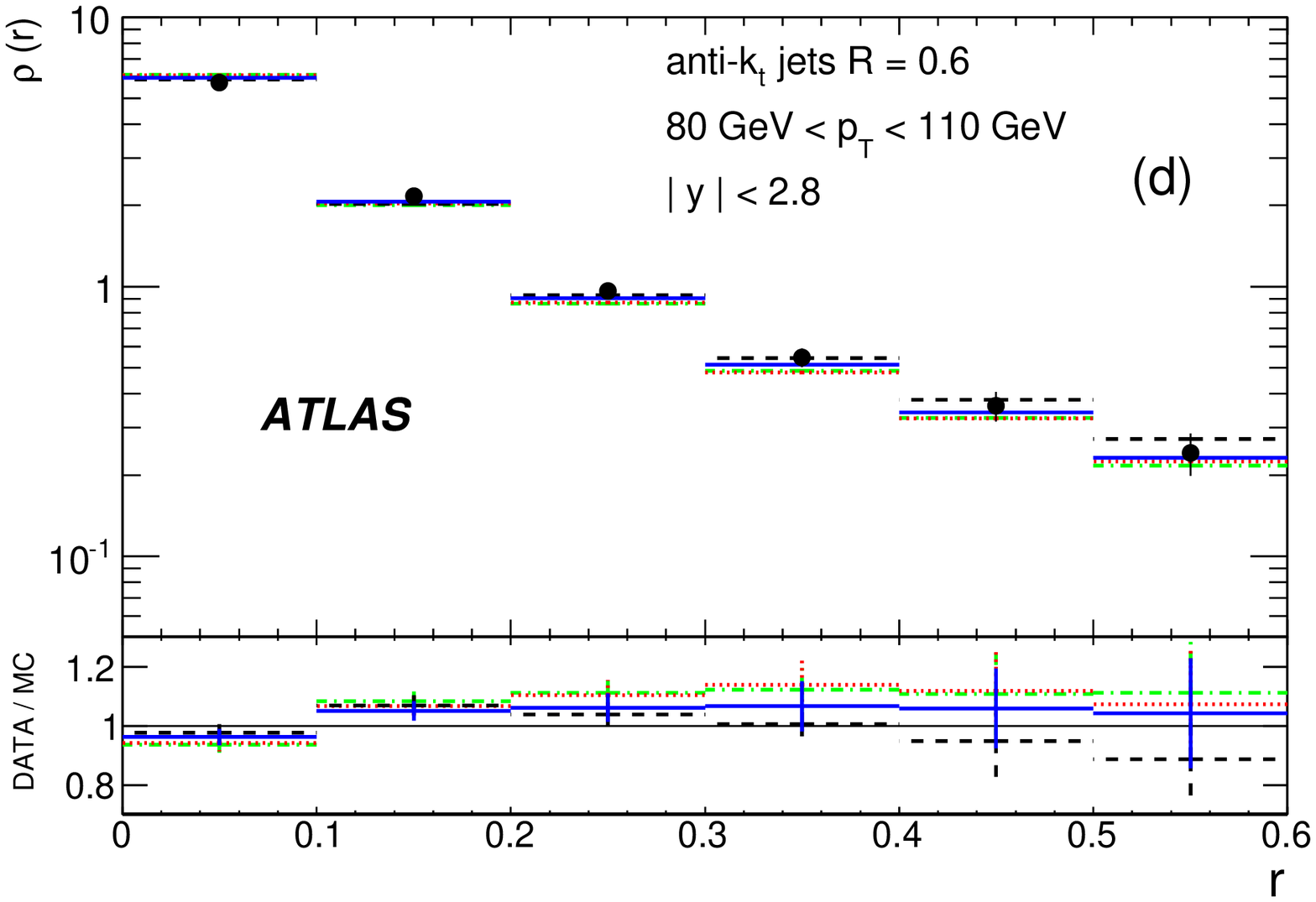}
}
\end{center}
\vspace{-0.7 cm}
\caption{\small
The measured differential jet shape, $\rho(r)$, in inclusive jet production for jets 
with $|\rapjet| < 2.8$ and $30 \ {\rm GeV} < \ptjet < 110  \ {\rm GeV}$   
is shown in different $\ptjet$ regions. Error bars indicate the statistical and systematic uncertainties added in quadrature.
The predictions of   PYTHIA-Perugia2010 (solid lines),   HERWIG++ (dashed lines),   ALPGEN interfaced with HERWIG and JIMMY (dotted lines), and 
  PYTHIA-MC09 (dashed-dotted lines) are shown for comparison.} 
\label{fig1}
\end{figure}


\begin{figure}[tbh]
\begin{center}
\mbox{
\includegraphics[width=0.5\textwidth,height=0.52\textwidth]{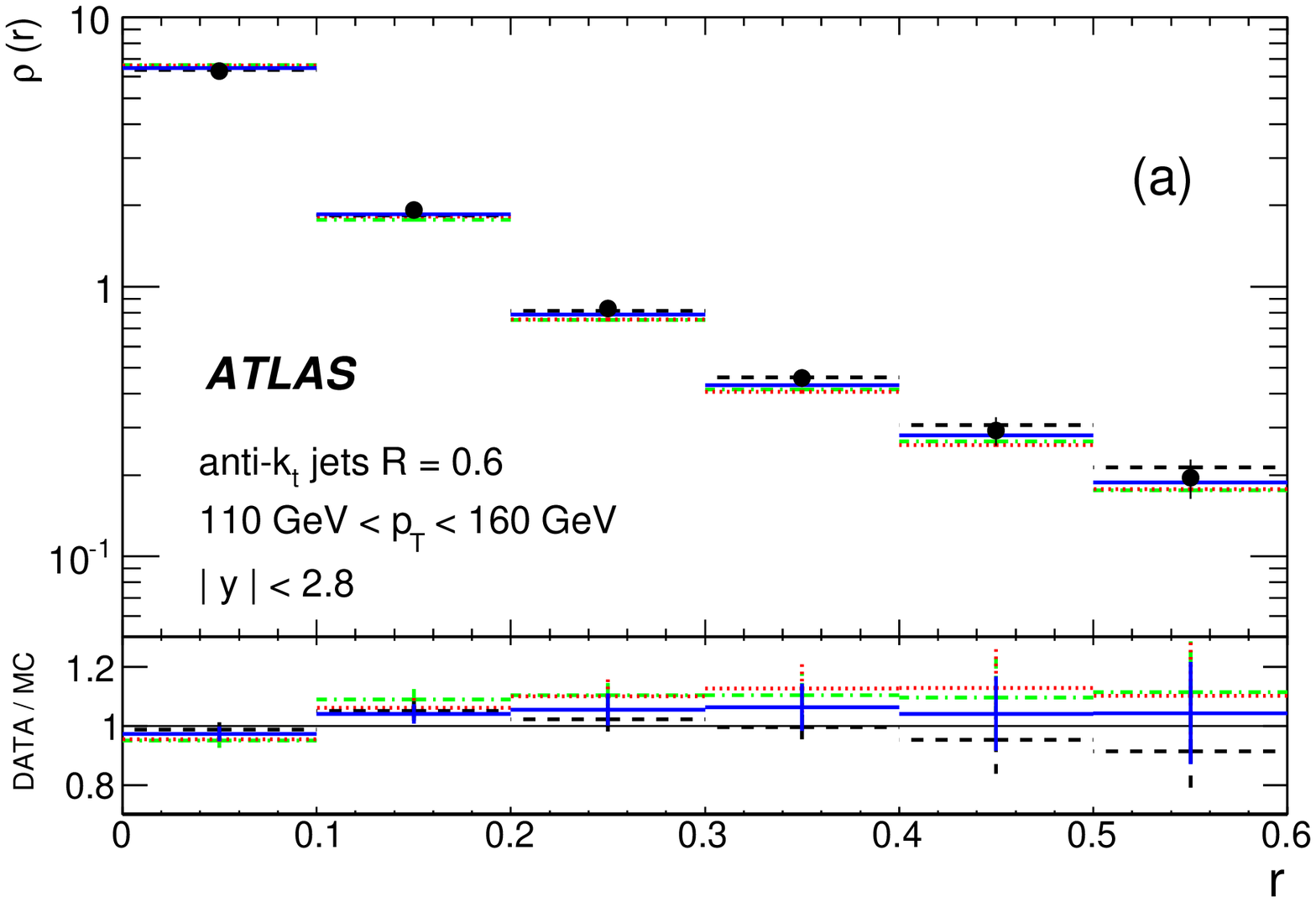} 
\includegraphics[width=0.5\textwidth,height=0.52\textwidth]{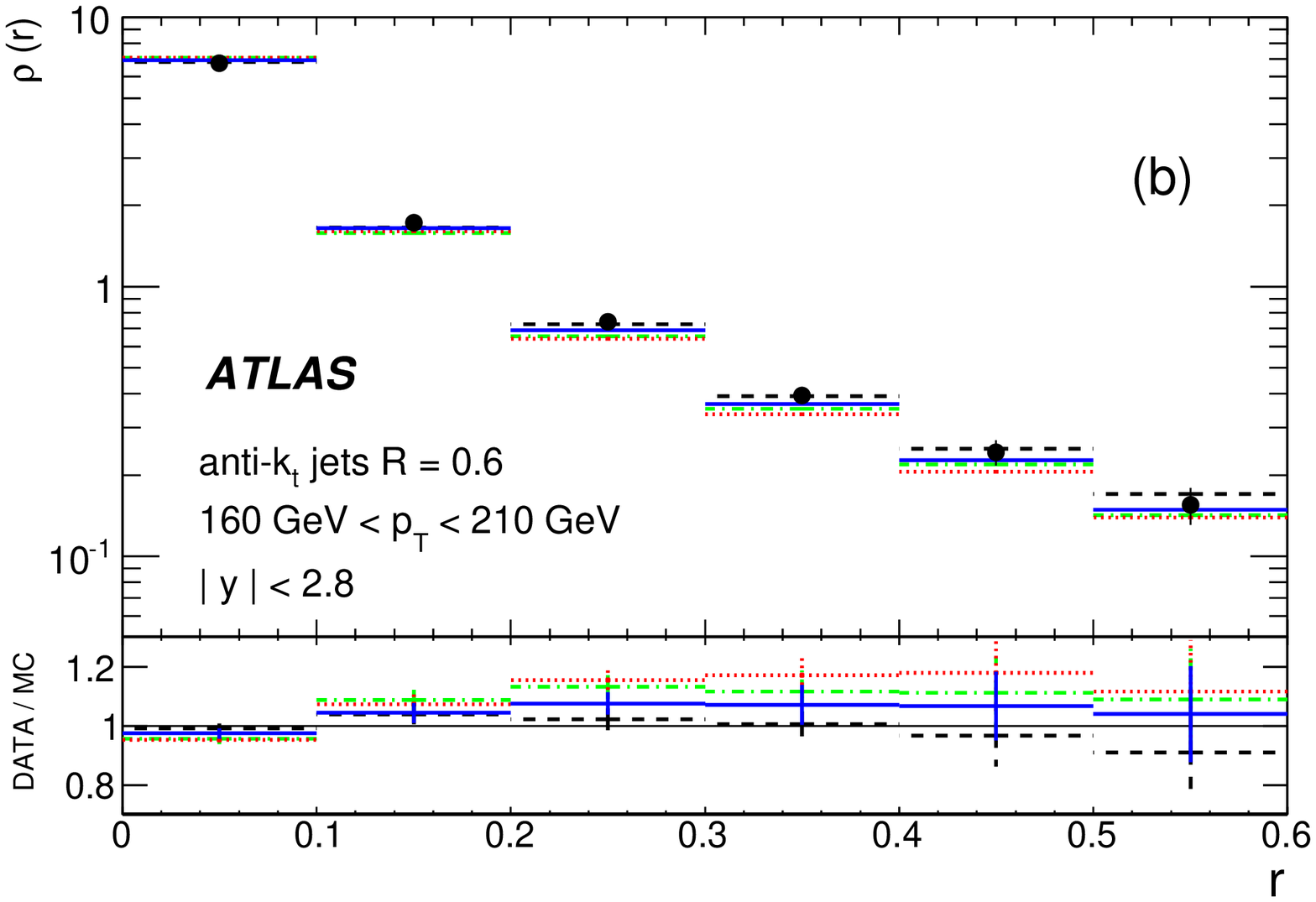}
}\vspace{-0.2cm}
\mbox{
\includegraphics[width=0.5\textwidth,height=0.52\textwidth]{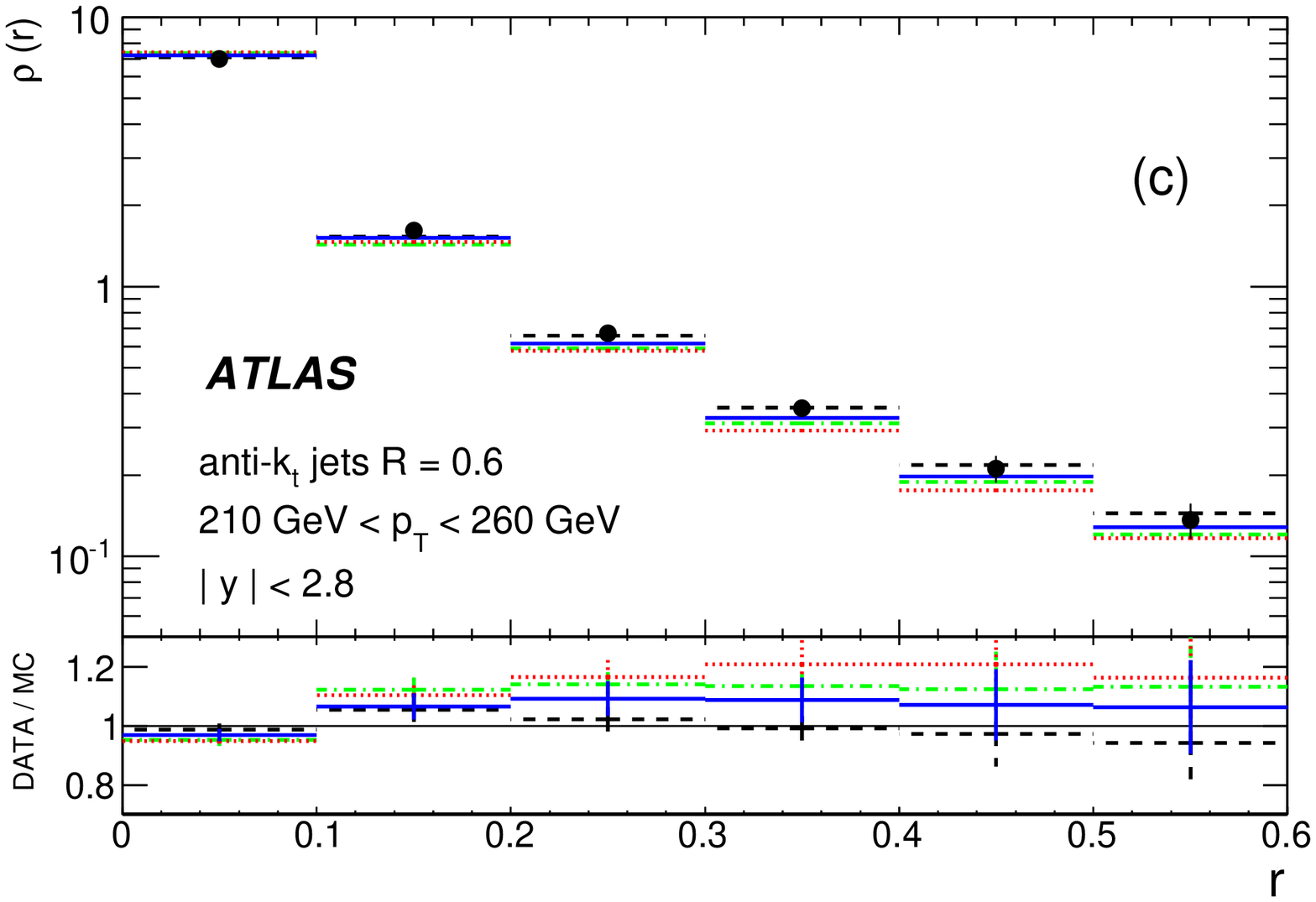} 
\includegraphics[width=0.5\textwidth,height=0.52\textwidth]{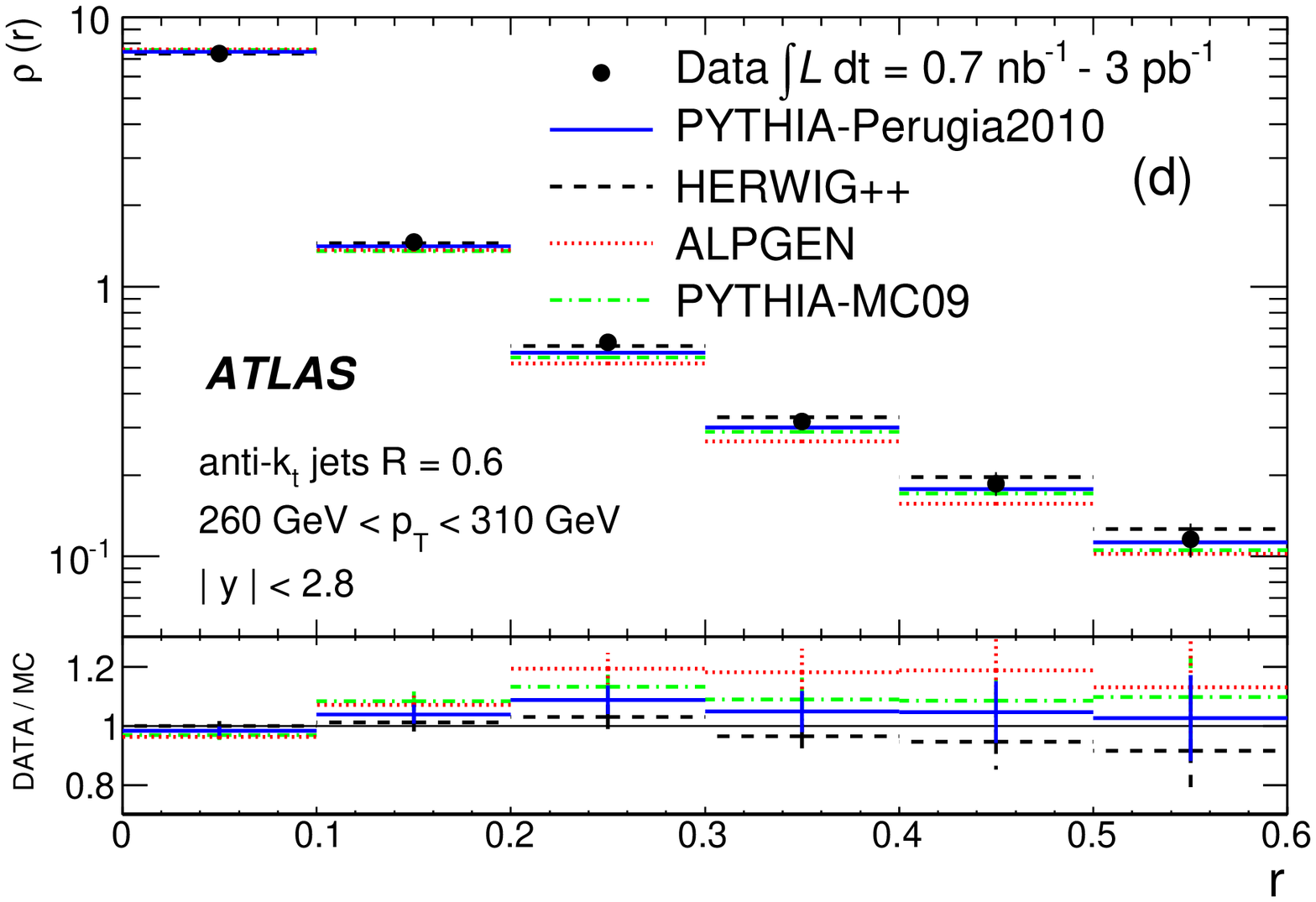} 
}
\end{center}
\vspace{-0.7 cm}
\caption{\small
The measured differential jet shape, $\rho(r)$, in inclusive jet production for jets 
with $|\rapjet| < 2.8$ and $110 \ {\rm GeV} < \ptjet < 310  \ {\rm GeV}$   
is shown in different $\ptjet$ regions. Error bars indicate the statistical and systematic uncertainties added in quadrature.
The predictions of   PYTHIA-Perugia2010 (solid lines),   HERWIG++ (dashed lines),   ALPGEN interfaced with HERWIG and JIMMY (dotted lines), and 
  PYTHIA-MC09 (dashed-dotted lines) are shown for comparison.} 
\label{fig1b}
\end{figure}

\clearpage


\begin{figure}[tbh]
\begin{center}
\mbox{
\includegraphics[width=0.5\textwidth,height=0.52\textwidth]{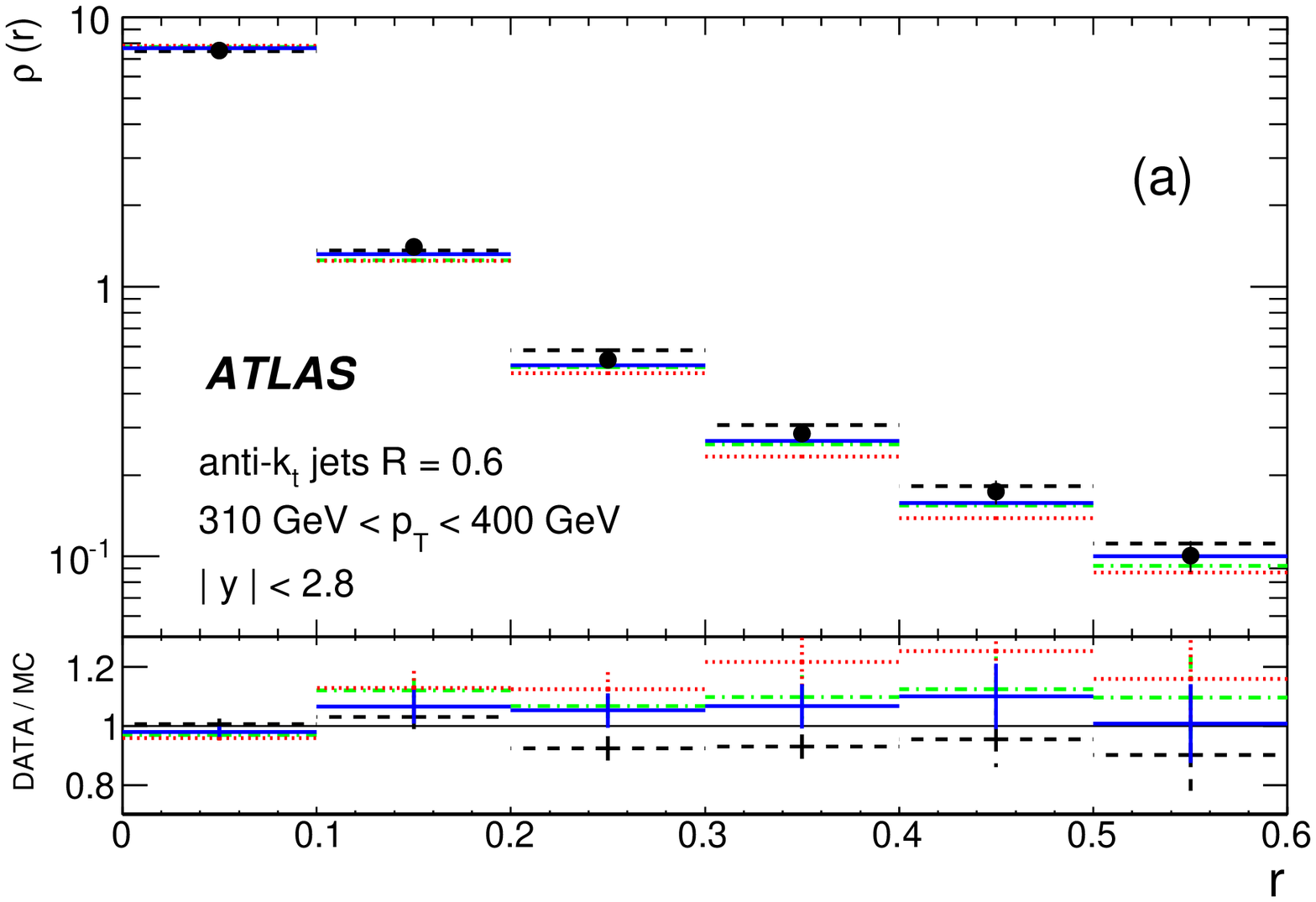} 
\includegraphics[width=0.5\textwidth,height=0.52\textwidth]{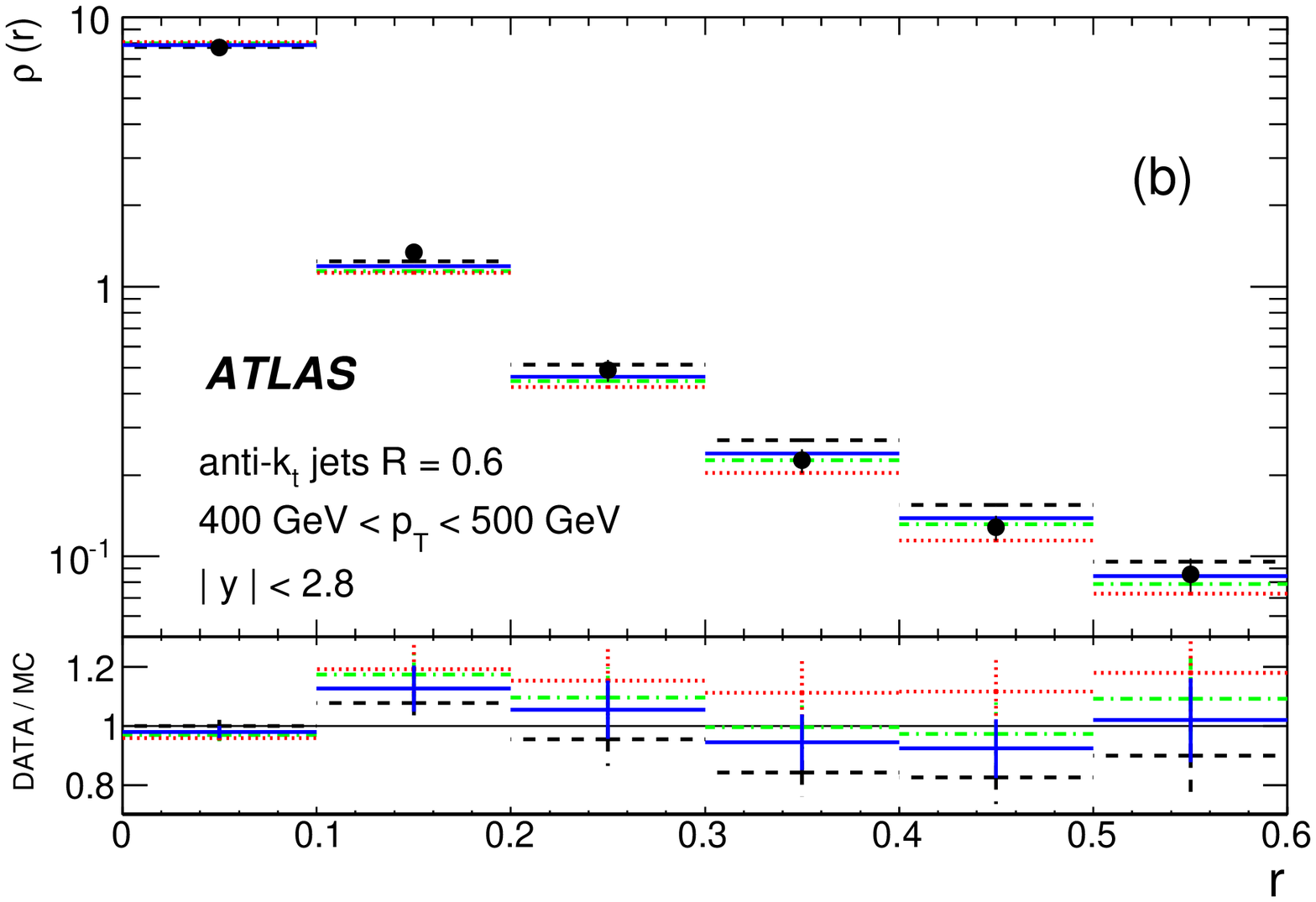}
}\vspace{-0.2 cm}
\mbox{
\includegraphics[width=0.5\textwidth,height=0.52\textwidth]{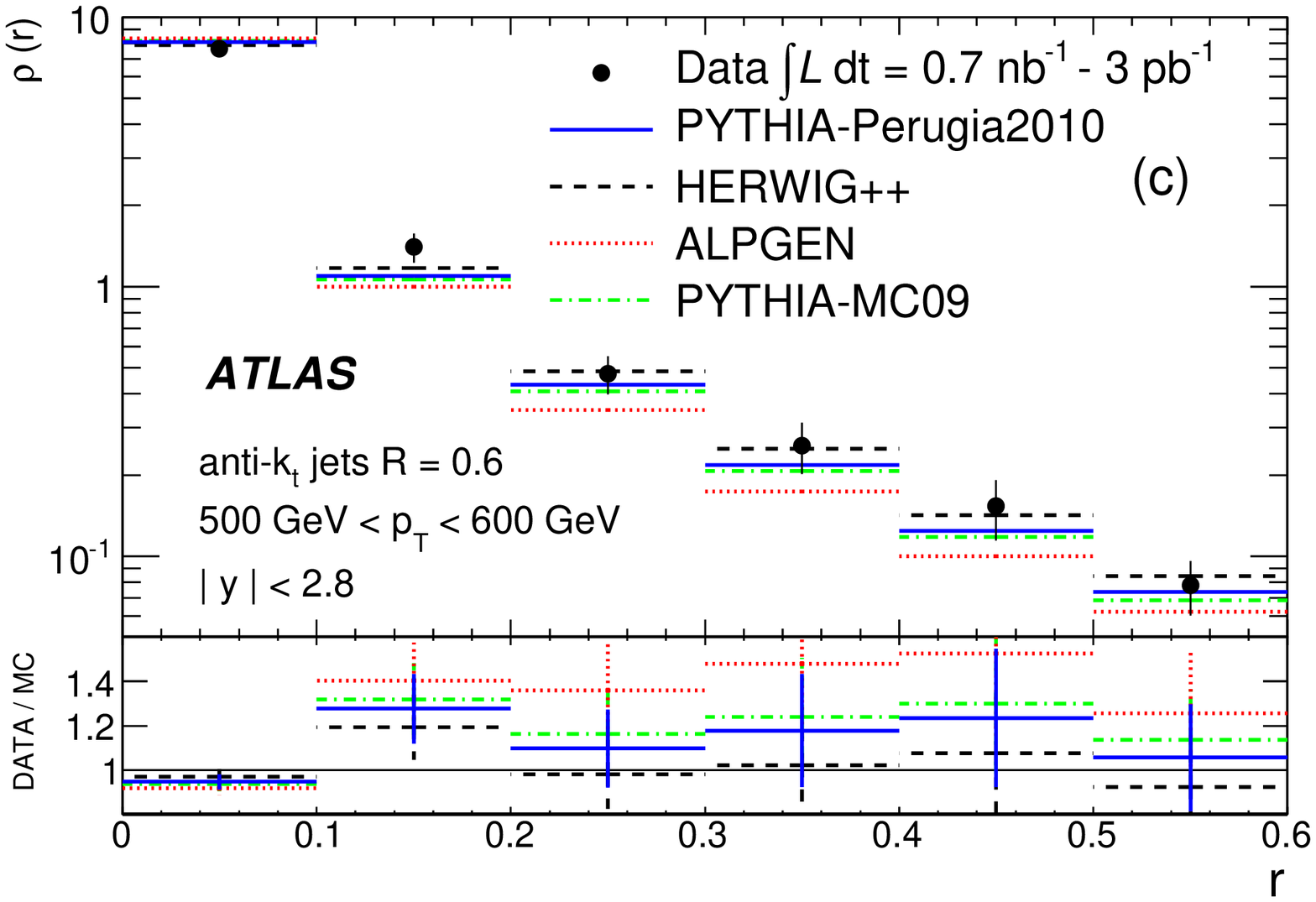}
}\vspace{-0.7 cm}
\caption{\small
The measured differential jet shape, $\rho(r)$, in inclusive jet production for jets 
with $|\rapjet| < 2.8$ and $310 \ {\rm GeV} < \ptjet < 600  \ {\rm GeV}$   
is shown in different $\ptjet$ regions. Error bars indicate the statistical and systematic uncertainties added in quadrature.
The predictions of   PYTHIA-Perugia2010 (solid lines),   HERWIG++ (dashed lines),   ALPGEN interfaced with HERWIG and JIMMY  (dotted lines), and   PYTHIA-MC09 (dashed-dotted lines) are shown for comparison.
} 
\label{fig2}
\end{center}
\end{figure}

\clearpage

\begin{figure}[tbh]
\begin{center}
\mbox{
\includegraphics[width=0.8\textwidth]{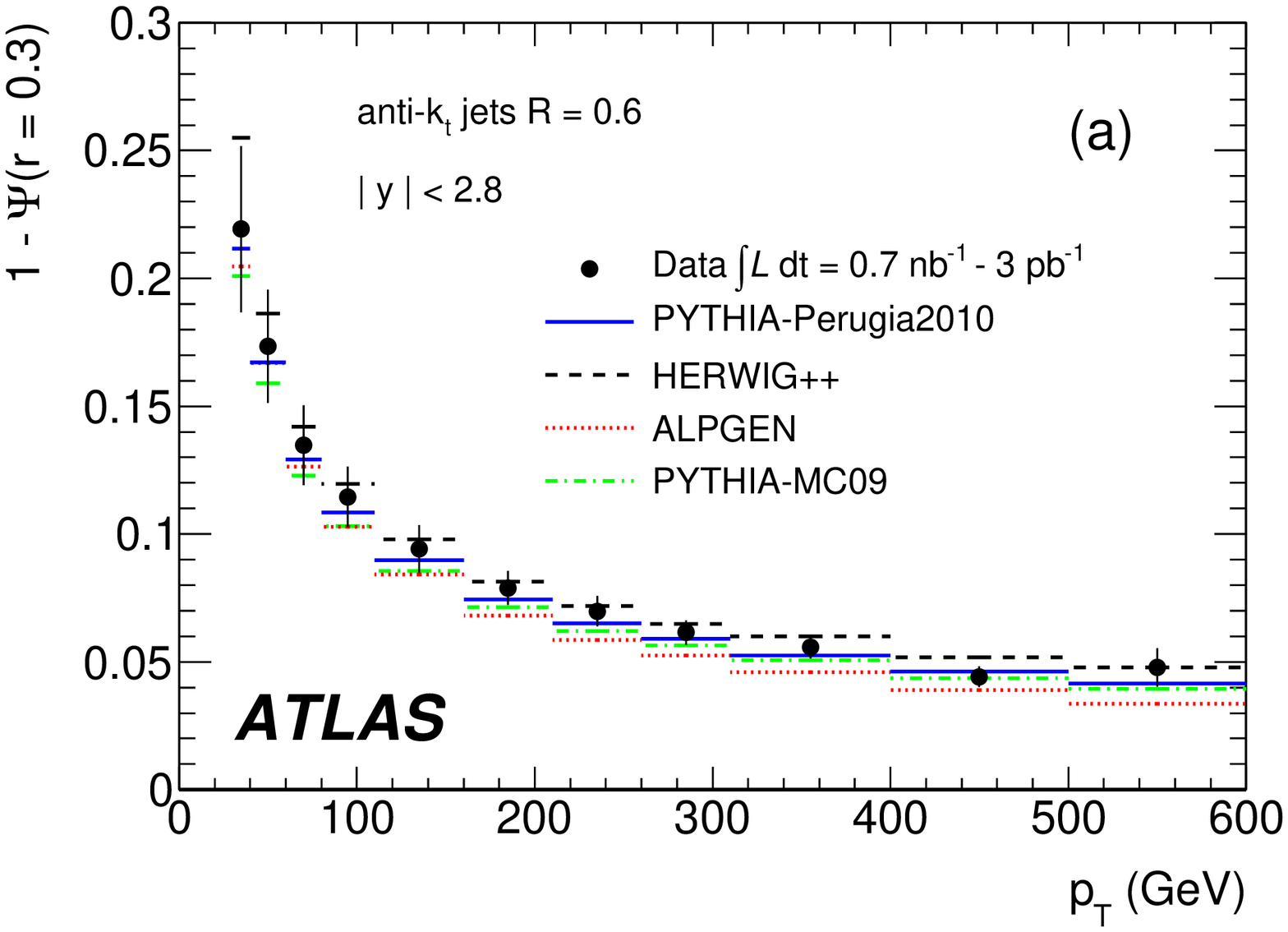} 
} \\
\mbox{
\includegraphics[width=0.8\textwidth]{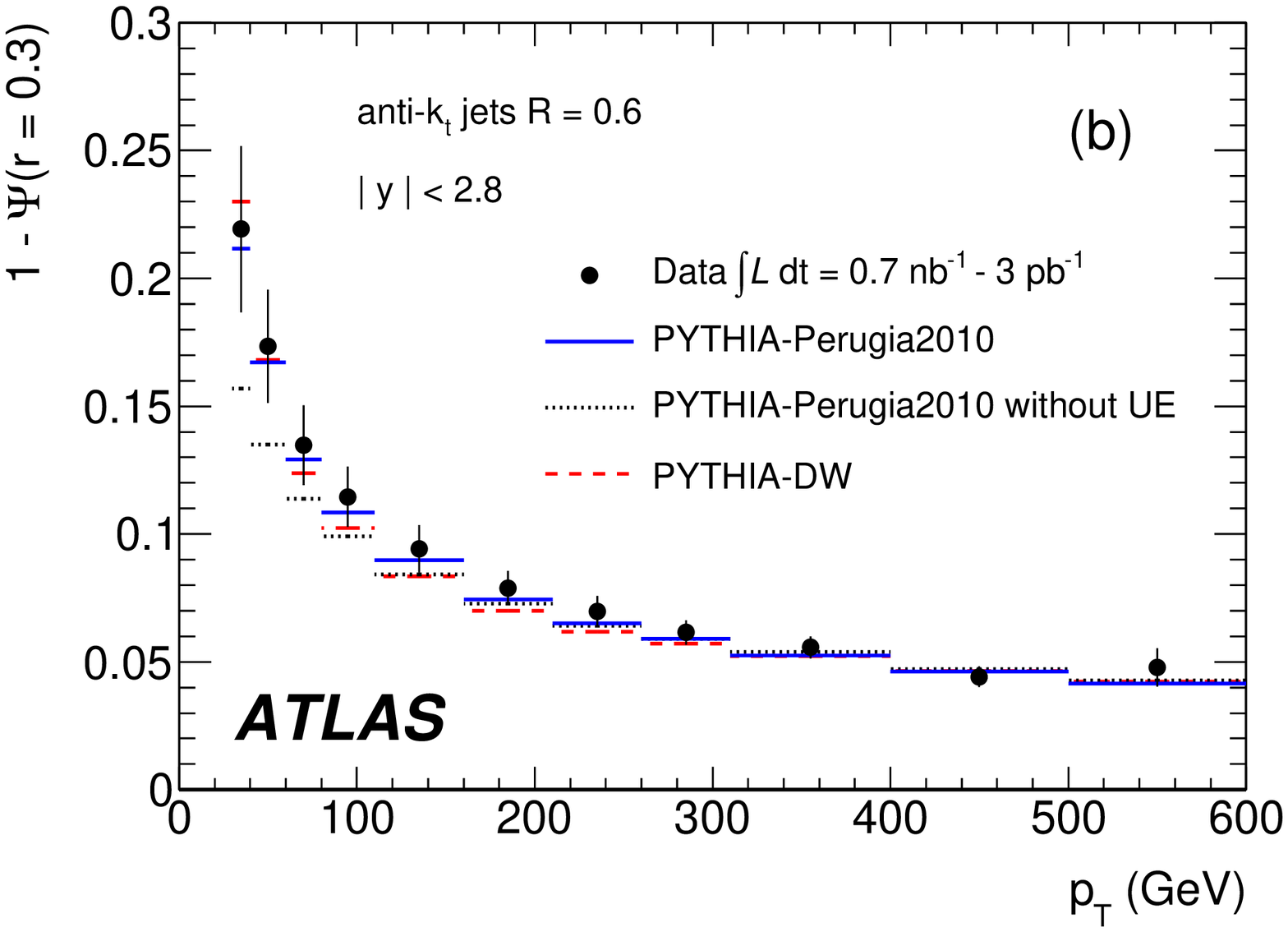} 
}
\end{center}
\vspace{-0.7 cm}
\caption{\small
The measured integrated jet shape, $1 - \Psi(r=0.3)$, as a function of $\ptjet$ 
for jets with $|\rapjet| < 2.8$ and $30 \ {\rm GeV} < \ptjet < 600 \ {\rm GeV}$.
Error bars indicate the statistical and systematic uncertainties added in quadrature.
The data are compared to the predictions of:   
(a) PYTHIA-Perugia2010 (solid lines),   HERWIG++ (dashed lines),   ALPGEN interfaced with HERWIG and JIMMY (dotted lines), and   PYTHIA-MC09 (dashed-dotted lines); (b) PYTHIA-Perugia2010 (solid lines),   PYTHIA-Perugia2010 without UE   (dotted lines), and   PYTHIA-DW (dashed lines).
} 
\label{fig3}
\end{figure}

\clearpage



\clearpage

\begin{figure}[tbh]
\begin{center}
\mbox{
\includegraphics[width=0.495\textwidth]{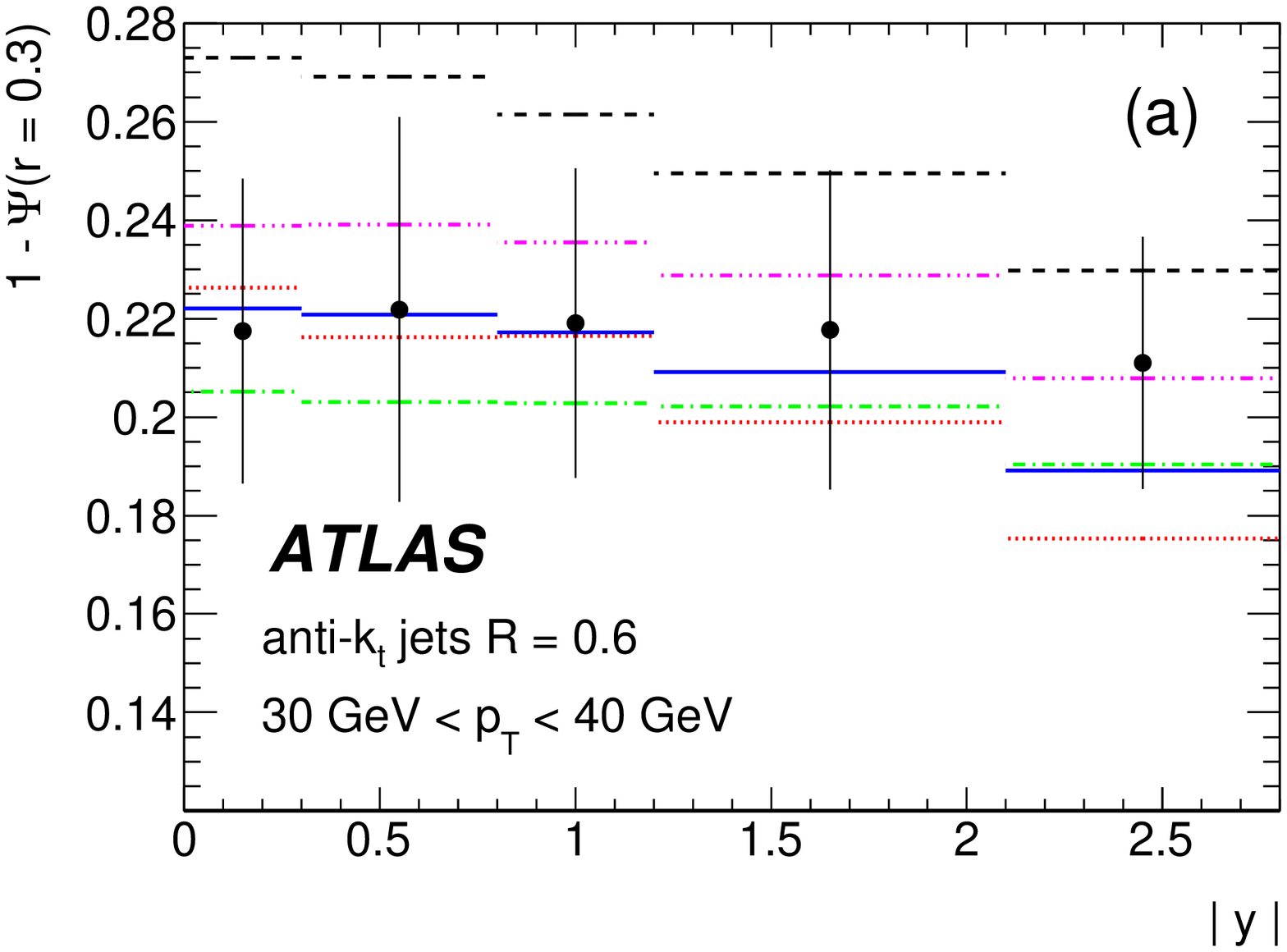} 
\includegraphics[width=0.495\textwidth]{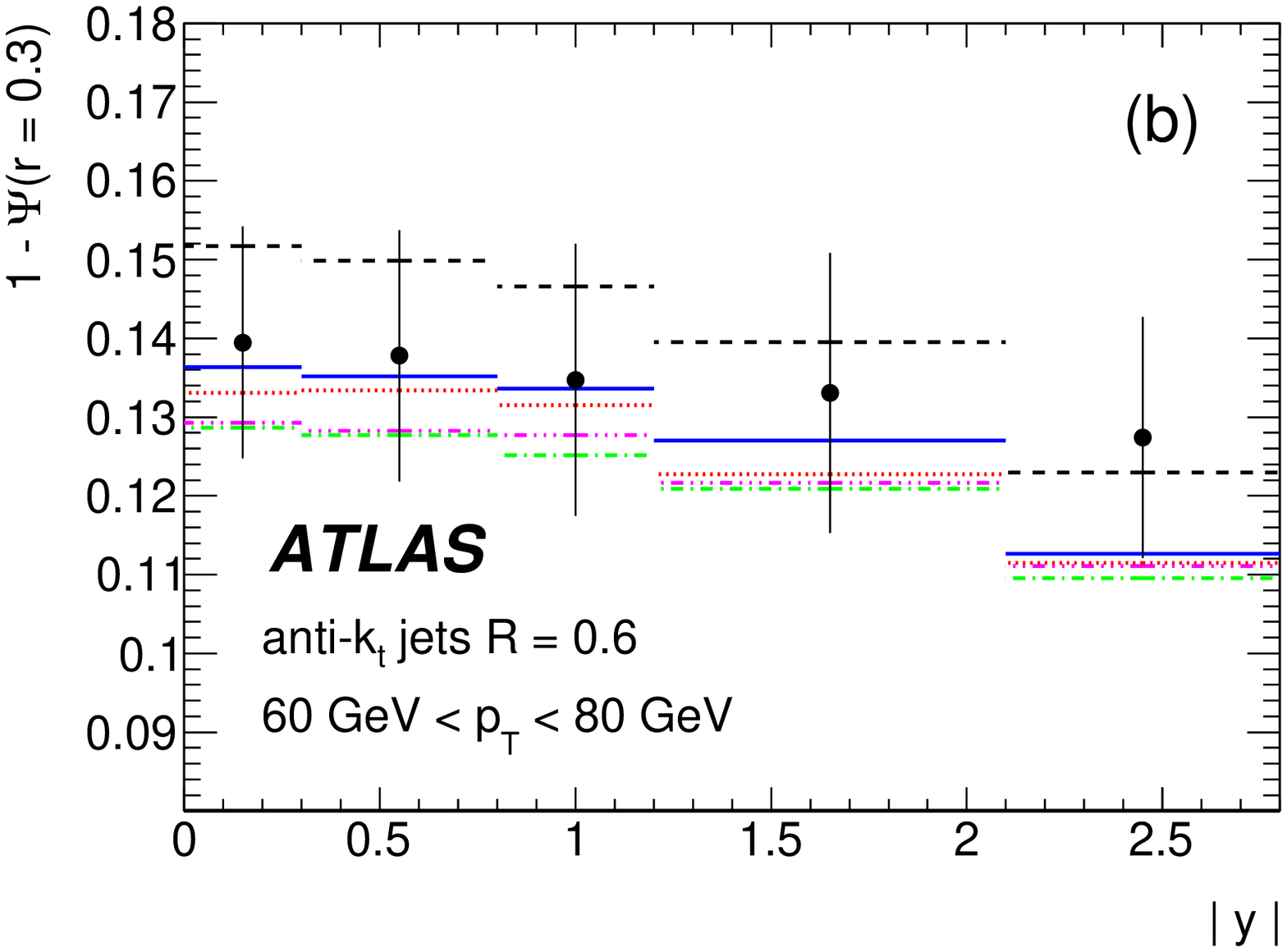} 
}
\mbox{
\includegraphics[width=0.495\textwidth]{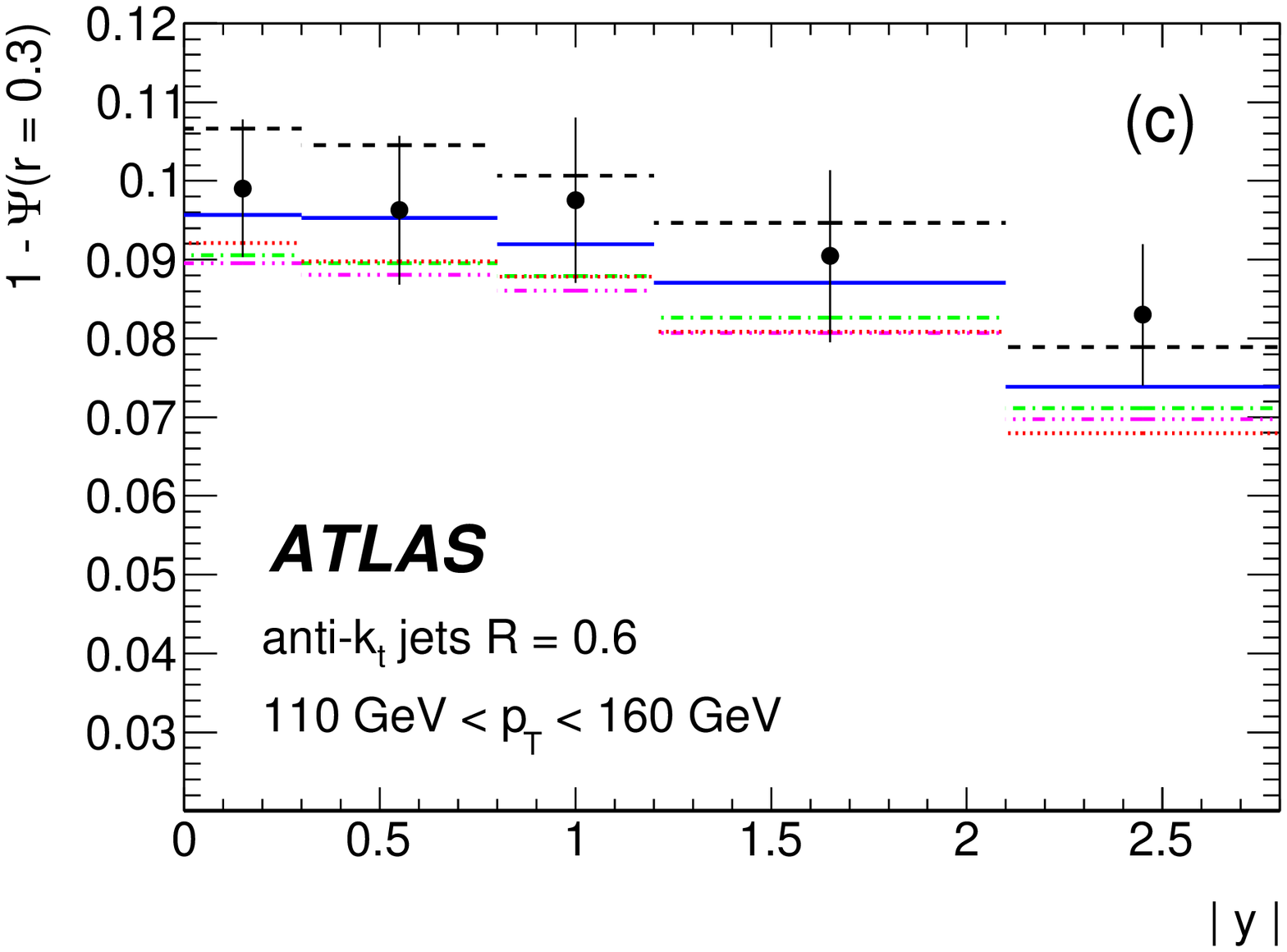}
\includegraphics[width=0.495\textwidth]{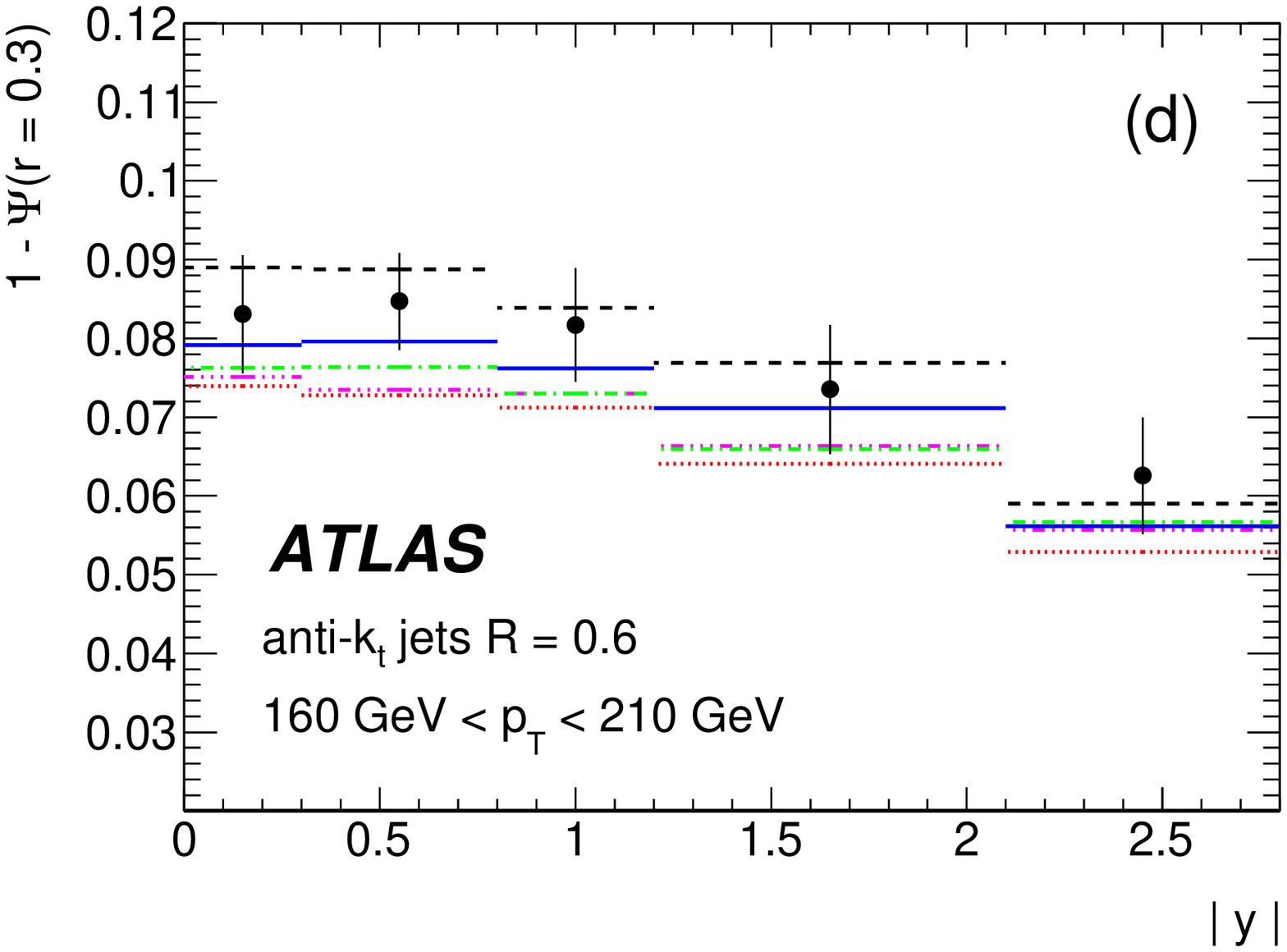} 
} 
\mbox{
\includegraphics[width=0.495\textwidth]{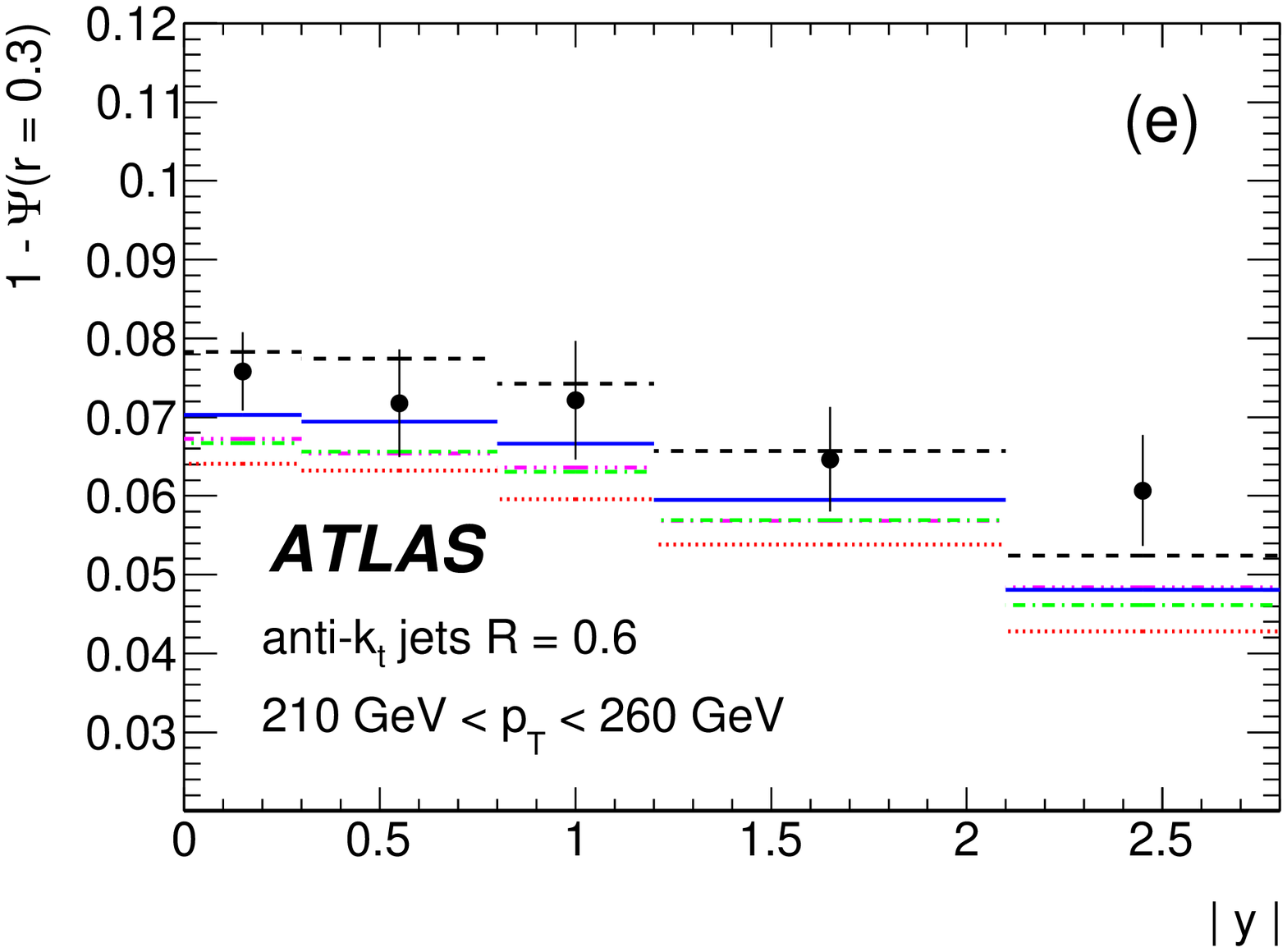} 
\includegraphics[width=0.495\textwidth]{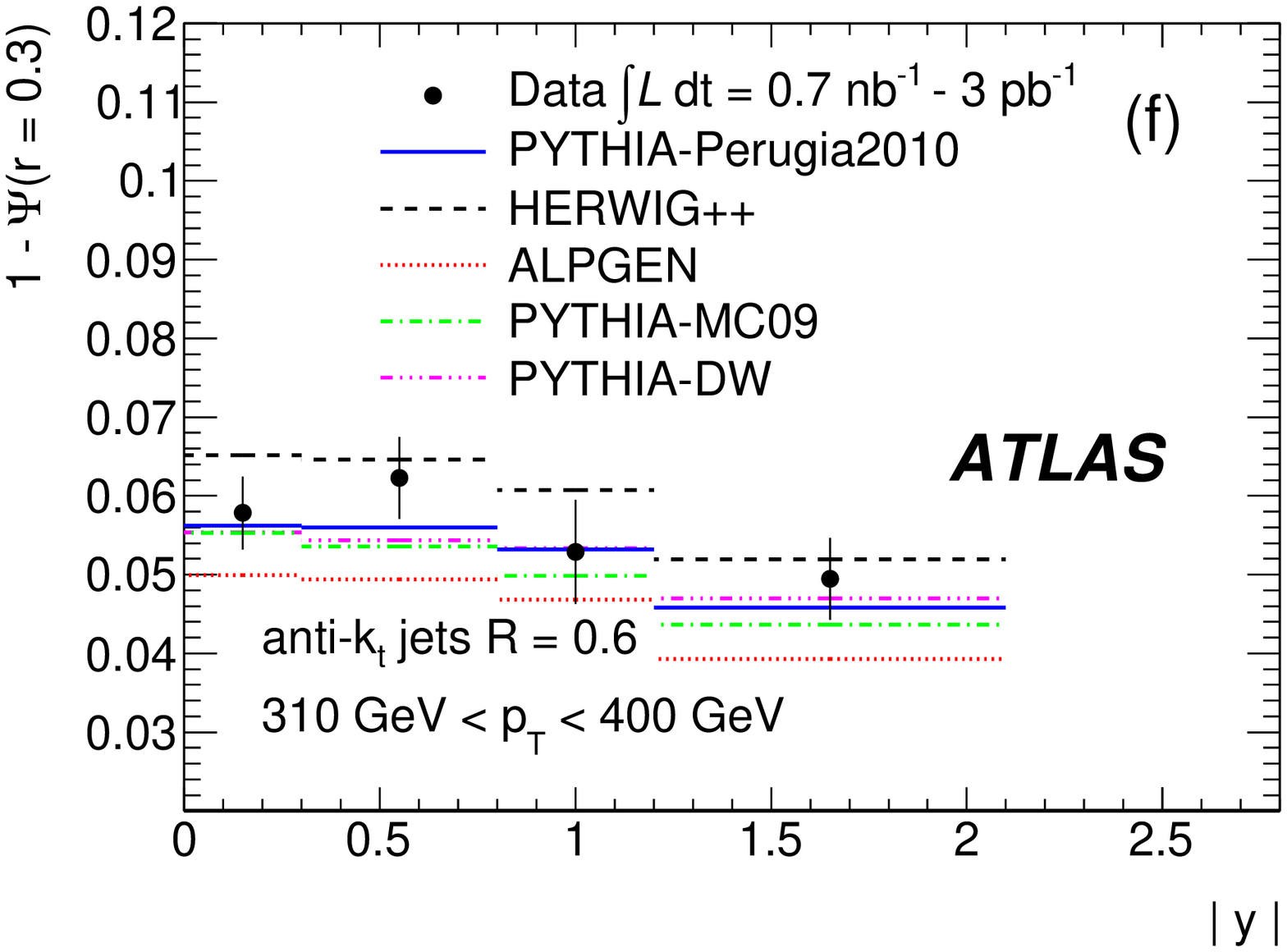}
}
\end{center}
\vspace{-0.7 cm}
\caption{\small
The measured integrated jet shape, $1 - \Psi(r=0.3)$, as a function of $|\rapjet|$ 
for jets with $|\rapjet| < 2.8$ and $30 \ {\rm GeV} < \ptjet < 400 \ {\rm GeV}$.
Error bars indicate the statistical and systematic uncertainties added in quadrature. 
The predictions of   PYTHIA-Perugia2010 (solid lines),   HERWIG++ (dashed lines),   ALPGEN interfaced with HERWIG and JIMMY (dotted lines), PYTHIA-MC09 (dashed-dotted lines), and PYTHIA-DW (dashed-dotted-dotted lines) are shown for comparison.
} 
\label{fig4}
\end{figure}

\clearpage


\begin{figure}[tbh]
\begin{center}
\mbox{
\includegraphics[width=0.95\textwidth]{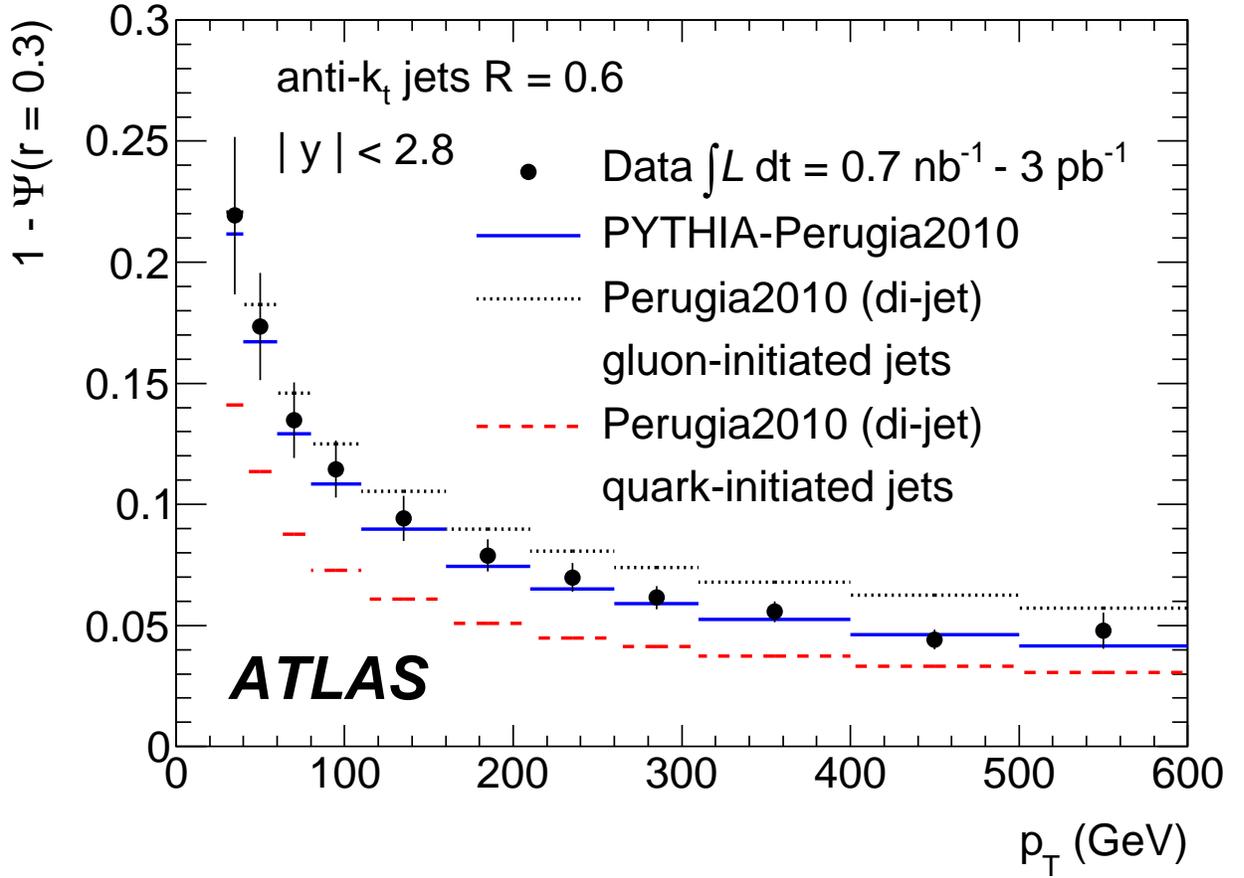}
}
\end{center}
\vspace{-0.7 cm}
\caption{\small
The measured integrated jet shape, $1 - \Psi(r=0.3)$, as a function of $\ptjet$ 
for jets with $|\rapjet| < 2.8$ and $30 \ {\rm GeV} < \ptjet < 600 \ {\rm GeV}$.
Error bars indicate the statistical and systematic uncertainties added in quadrature. 
The predictions of   PYTHIA-Perugia2010 (solid line) are shown for comparison, together with 
the prediction separately for  quark-initiated  (dashed lines) and
gluon-initiated jets (dotted lines) in dijet events.
} 
\label{fig5}
\end{figure}

\begin{figure}[tbh]
\begin{center}
\mbox{
\includegraphics[width=0.495\textwidth]{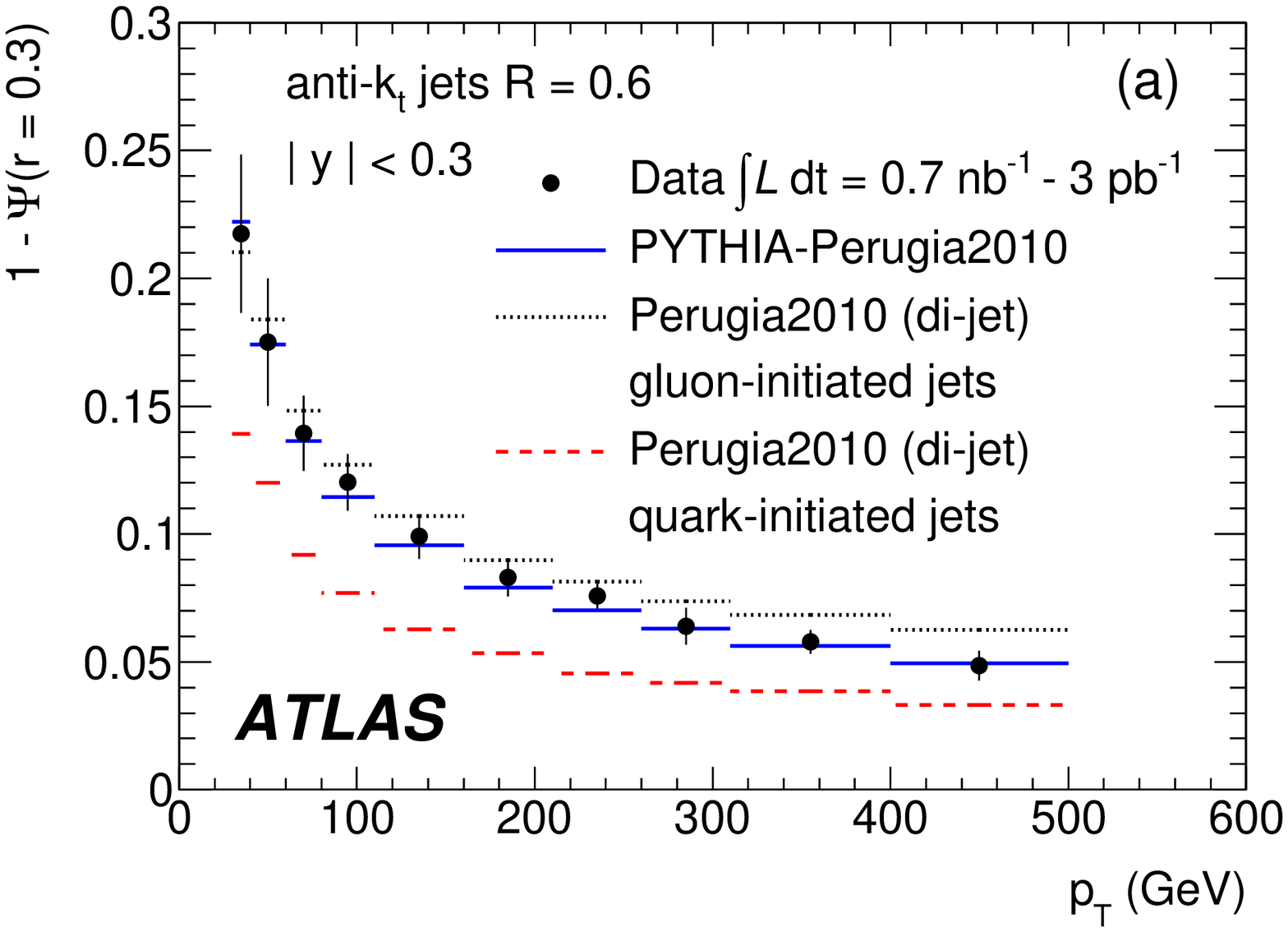}
\includegraphics[width=0.495\textwidth]{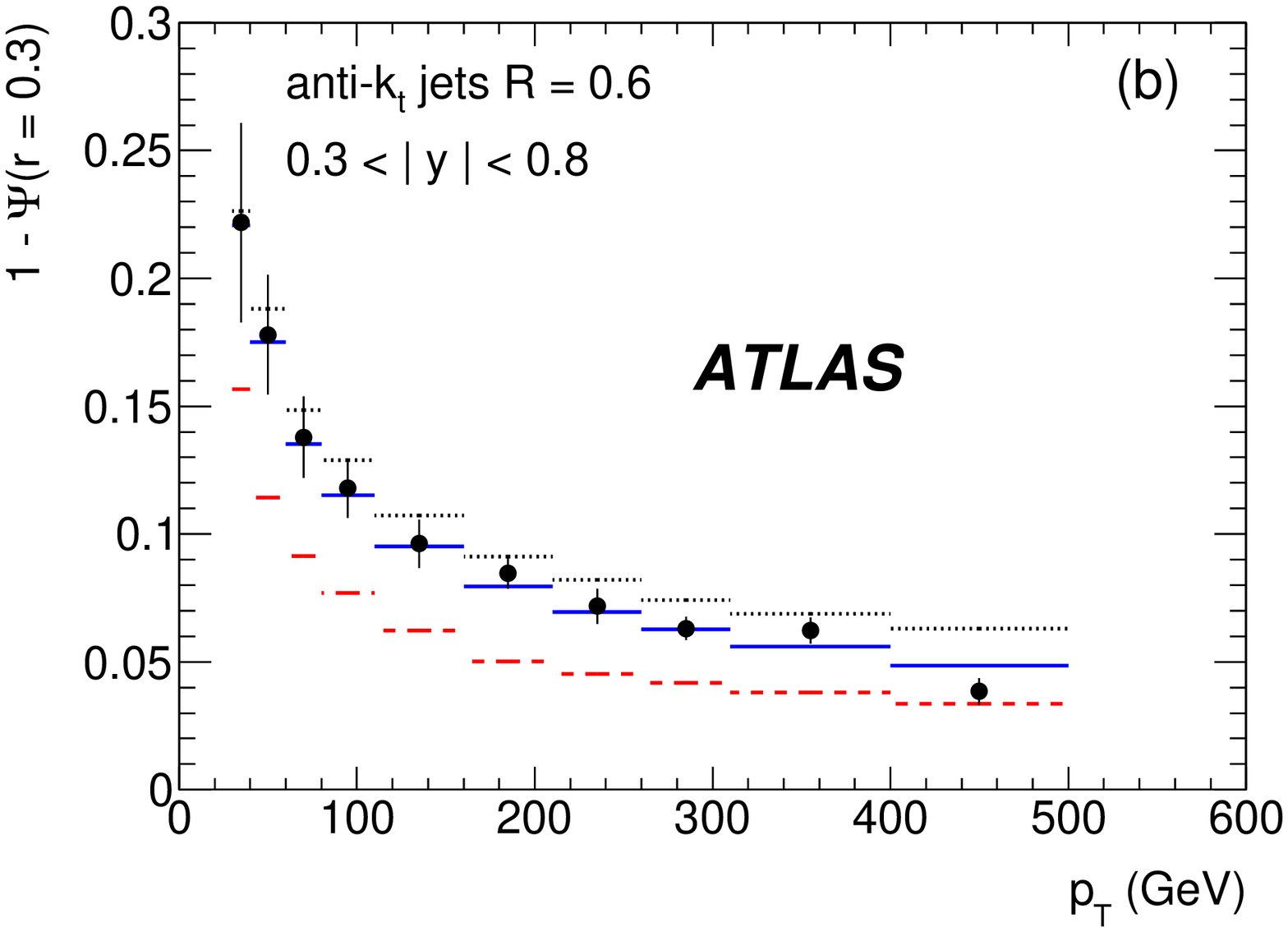}
}
\mbox{
\includegraphics[width=0.495\textwidth]{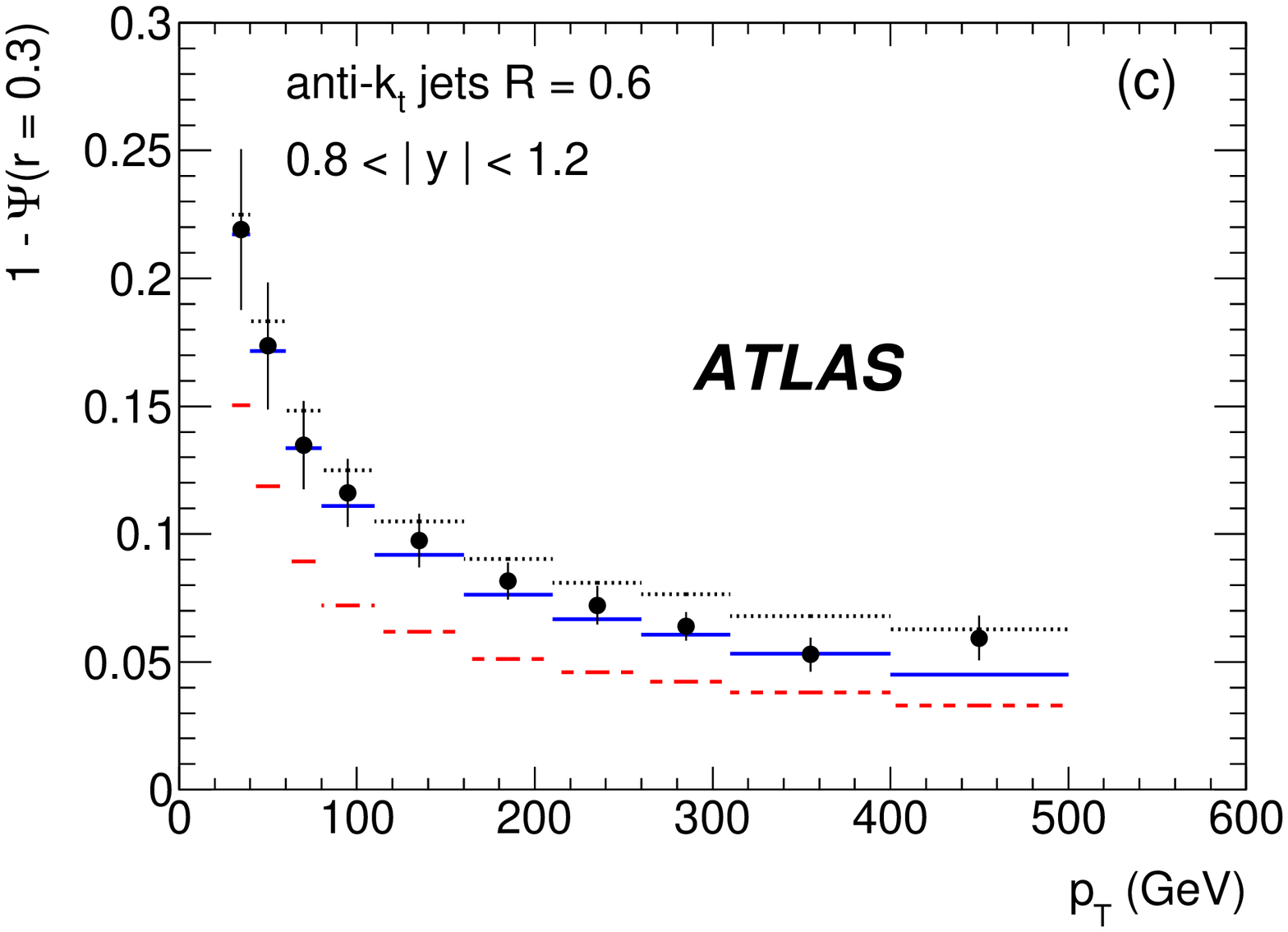} 
\includegraphics[width=0.495\textwidth]{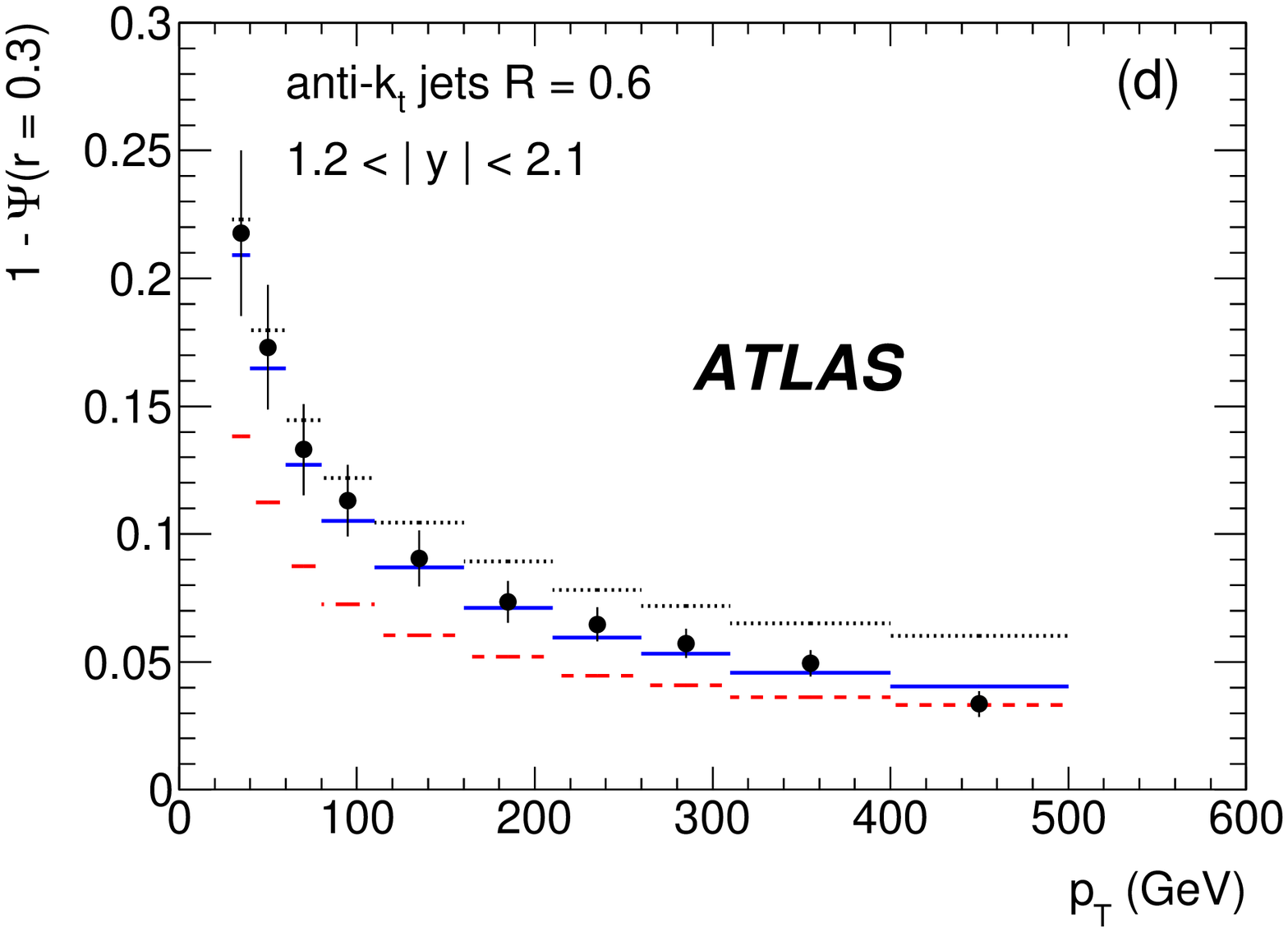}
}
\mbox{
\includegraphics[width=0.495\textwidth]{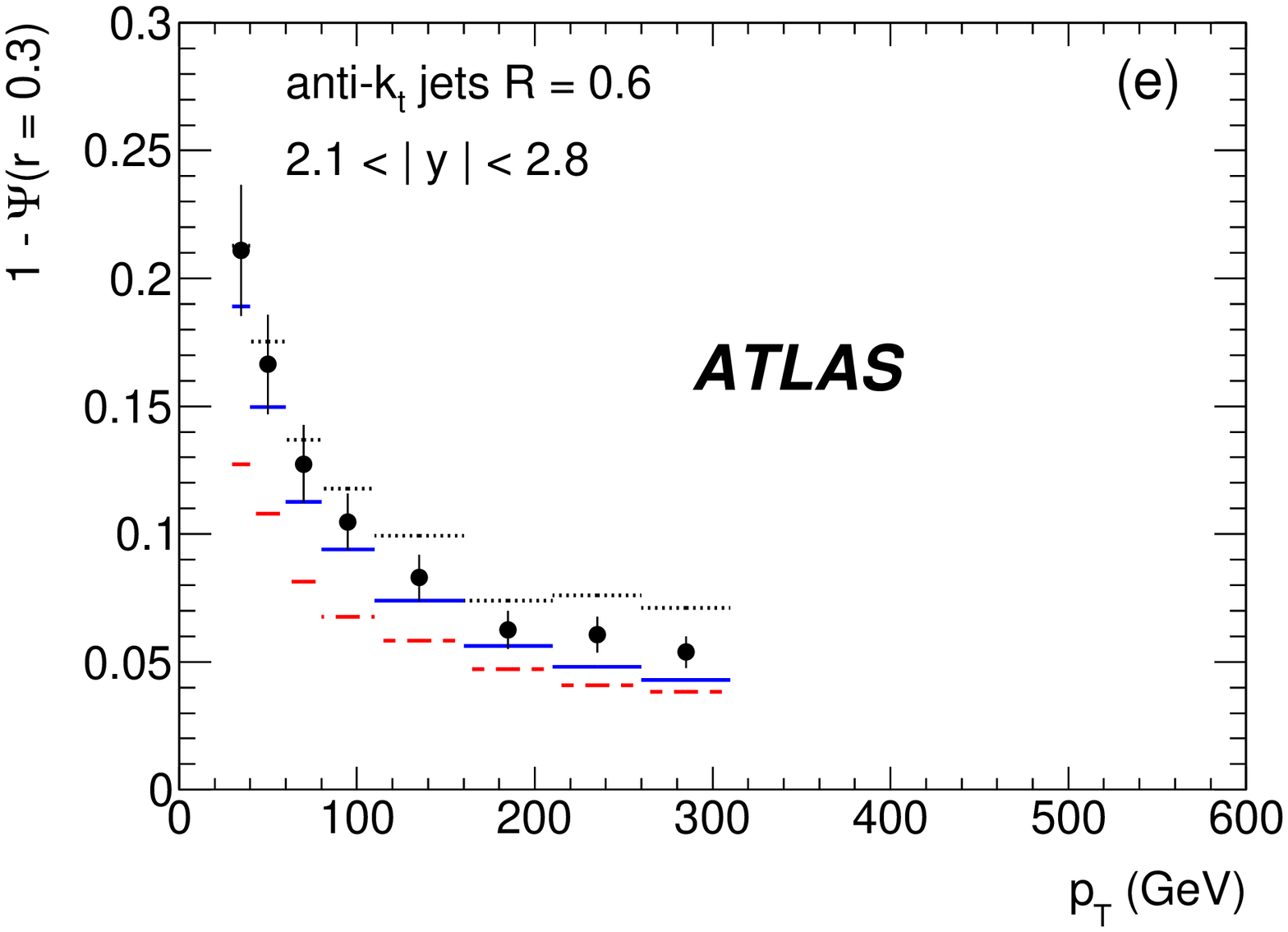}
}
\end{center}
\vspace{-0.7 cm}
\caption{\small
The measured integrated jet shape, $1 - \Psi(r=0.3)$, as a function of $\ptjet$ in different jet rapidity regions 
for jets with $|\rapjet| < 2.8$ and $30 \ {\rm GeV} < \ptjet < 500 \ {\rm GeV}$.
Error bars indicate the statistical and systematic uncertainties added in quadrature. 
The predictions of   PYTHIA-Perugia2010 (solid line) are shown for comparison, together with 
the prediction separately for  quark-initiated  (dashed lines) and
gluon-initiated jets (dotted lines) in dijet events.
} 
\label{fig6}
\end{figure}

\clearpage



\appendix
\section{Data Points and Correlation of Systematic Uncertainties}

Data for differential and integrated measurements are collected in Tables~2 to 
6, which include a detailed description of the contributions from the different
sources of systematic uncertainty, as discussed in Section~8.

A $\chi^2$ test is performed  to the data points in Tables~5 and 6 
with respect to a given MC prediction, separately in each rapidity region. 
The systematic uncertainties are considered independent and fully correlated across $\ptjet$ bins, 
and the test is carried out according to the formula

\begin{equation}
\chi^2 = \sum_{j=1}^{\ptjet \  bins} \frac{[{  d}_j - {  mc}_j(\bar{s})]^2}{[\delta{  d}_j]^2 + [\delta{  mc}_j(\bar{s})]^2
 } 
+ \sum_{i=1}^{5} [s_i]^2 \ , 
\end{equation}
\noindent
where ${  d}_j$ is the measured data point $j$,   ${  mc}_j(\bar{s})$ is 
the corresponding MC prediction, and  $\bar{s}$  
denotes the vector of  standard deviations, $s_i$, for the different independent sources of 
systematic uncertainty.
For each rapidity region considered,  the sums above  run over the total number of data points 
in $\ptjet$ and five independent sources of systematic uncertainty, and the $\chi^2$ is minimized with respect to $\bar{s}$.
Correlations among systematic uncertainties  
are taken into account in ${  mc}_j(\bar{s})$.  The $\chi^2$ results for the different 
MC predictions are collected in Table~7,  and indicate that PYTHIA-Perugia2010 provides the overall 
best description of the data. 


\begin{sidewaystable}
\begin{tiny}
\begin{center}
\begin{tabular}{|c|c||c|c|c|c|c|c|} \hline
\multicolumn{8}{|c|}{{\large{$\rho(r) \ (0 < |\rapjet|< 2.8)$}}} \\ \hline\hline\hline
\multicolumn{8}{|c|}{$30$ GeV $< \ptjet < 40$ GeV} \\ \hline
$r$& $\rho \pm \ \rm (stat.) \pm \ \rm (syst.)$ &cluster e-scale & shower model &jet e-scale & resolution & correction & non-closure\\ \hline

0.05 & 3.486 $\pm$ 0.011 $\pm$ 0.331 & $\mp$ 0.202 & $\pm$ 0.024 & $\mp$ 0.168 & $\pm$ 0.102 & $\pm$ 0.169 & $\pm$ 0.035 \\
0.15 & 2.787 $\pm$ 0.009 $\pm$ 0.093 & $\mp$ 0.034 & $\mp$ 0.024 & $\pm$ 0.022 & $\pm$ 0.015 & $\pm$ 0.074 & $\pm$ 0.028 \\
0.25 & 1.550 $\pm$ 0.006 $\pm$ 0.102 & $\pm$ 0.048 & $\mp$ 0.005 & $\pm$ 0.076 & $\pm$ 0.041 & $\mp$ 0.019 & $\pm$ 0.015 \\
0.35 & 0.995 $\pm$ 0.004 $\pm$ 0.112 & $\pm$ 0.072 & $\mp$ 0.007 & $\pm$ 0.058 & $\pm$ 0.032 & $\mp$ 0.054 & $\pm$ 0.010 \\
0.45 & 0.748 $\pm$ 0.003 $\pm$ 0.117 & $\pm$ 0.088 & $\pm$ 0.005 & $\pm$ 0.044 & $\pm$ 0.022 & $\mp$ 0.059 & $\pm$ 0.007 \\
0.55 & 0.455 $\pm$ 0.002 $\pm$ 0.105 & $\pm$ 0.082 & $\pm$ 0.002 & $\pm$ 0.019 & $\pm$ 0.009 & $\mp$ 0.062 & $\pm$ 0.005 \\ \hline\hline
\multicolumn{8}{|c|}{$40$ GeV $< \ptjet < 60$ GeV} \\ \hline
$r$ & $\rho \pm \ \rm (stat.) \pm \ \rm (syst.)$ &cluster e-scale & shower model &jet e-scale & resolution & correction & non-closure\\ \hline
0.05 & 4.350 $\pm$ 0.018 $\pm$ 0.250 & $\mp$ 0.180 & $\pm$ 0.043 & $\mp$ 0.133 & $\pm$ 0.058 & $\pm$ 0.075 & $\pm$ 0.043 \\
0.15 & 2.631 $\pm$ 0.014 $\pm$ 0.072 & $\mp$ 0.026 & $\mp$ 0.038 & $\pm$ 0.032 & $\pm$ 0.001 & $\pm$ 0.036 & $\pm$ 0.026 \\
0.25 & 1.292 $\pm$ 0.009 $\pm$ 0.068 & $\pm$ 0.040 & $\mp$ 0.021 & $\pm$ 0.043 & $\pm$ 0.020 & $\mp$ 0.011 & $\pm$ 0.013 \\
0.35 & 0.797 $\pm$ 0.006 $\pm$ 0.081 & $\pm$ 0.064 & $\pm$ 0.004 & $\pm$ 0.034 & $\pm$ 0.024 & $\mp$ 0.026 & $\pm$ 0.008 \\
0.45 & 0.567 $\pm$ 0.004 $\pm$ 0.084 & $\pm$ 0.069 & $\mp$ 0.001 & $\pm$ 0.030 & $\pm$ 0.011 & $\mp$ 0.035 & $\pm$ 0.006 \\
0.55 & 0.372 $\pm$ 0.003 $\pm$ 0.075 & $\pm$ 0.065 & $\pm$ 0.004 & $\pm$ 0.014 & $\pm$ 0.009 & $\mp$ 0.034 & $\pm$ 0.004 \\ \hline\hline
\multicolumn{8}{|c|}{$60$ GeV $< \ptjet < 80$ GeV} \\ \hline
$r$ & $\rho \pm \ \rm (stat.) \pm \ \rm (syst.)$ &cluster e-scale & shower model &jet e-scale & resolution & correction & non-closure\\ \hline
0.05 & 5.193 $\pm$ 0.011 $\pm$ 0.210 & $\mp$ 0.149 & $\pm$ 0.076 & $\mp$ 0.087 & $\pm$ 0.048 & $\pm$ 0.058 & $\pm$ 0.052 \\
0.15 & 2.383 $\pm$ 0.007 $\pm$ 0.093 & $\mp$ 0.015 & $\mp$ 0.069 & $\pm$ 0.038 & $\pm$ 0.020 & $\pm$ 0.034 & $\pm$ 0.024 \\
0.25 & 1.074 $\pm$ 0.005 $\pm$ 0.058 & $\pm$ 0.033 & $\mp$ 0.026 & $\pm$ 0.020 & $\pm$ 0.014 & $\mp$ 0.029 & $\pm$ 0.011 \\
0.35 & 0.626 $\pm$ 0.003 $\pm$ 0.056 & $\pm$ 0.050 & $\pm$ 0.003 & $\pm$ 0.017 & $\pm$ 0.007 & $\mp$ 0.015 & $\pm$ 0.006 \\
0.45 & 0.437 $\pm$ 0.002 $\pm$ 0.060 & $\pm$ 0.055 & $\pm$ 0.006 & $\pm$ 0.014 & $\pm$ 0.010 & $\mp$ 0.015 & $\pm$ 0.004 \\
0.55 & 0.288 $\pm$ 0.001 $\pm$ 0.056 & $\pm$ 0.049 & $\pm$ 0.002 & $\pm$ 0.009 & $\pm$ 0.001 & $\mp$ 0.026 & $\pm$ 0.003 \\ \hline\hline
\multicolumn{8}{|c|}{$80$ GeV $< \ptjet < 110$ GeV} \\ \hline
$r$ & $\rho \pm \ \rm (stat.) \pm \ \rm (syst.)$ &cluster e-scale & shower model &jet e-scale & resolution & correction & non-closure\\ \hline
0.05 & 5.719 $\pm$ 0.010 $\pm$ 0.169 & $\mp$ 0.122 & $\pm$ 0.075 & $\mp$ 0.061 & $\pm$ 0.014 & $\pm$ 0.031 & $\pm$ 0.057 \\
0.15 & 2.166 $\pm$ 0.006 $\pm$ 0.068 & $\mp$ 0.009 & $\mp$ 0.054 & $\pm$ 0.026 & $\pm$ 0.001 & $\pm$ 0.021 & $\pm$ 0.022 \\
0.25 & 0.962 $\pm$ 0.004 $\pm$ 0.043 & $\pm$ 0.026 & $\mp$ 0.023 & $\pm$ 0.014 & $\pm$ 0.007 & $\mp$ 0.019 & $\pm$ 0.010 \\
0.35 & 0.547 $\pm$ 0.002 $\pm$ 0.044 & $\pm$ 0.041 & $\mp$ 0.007 & $\pm$ 0.013 & $\pm$ 0.002 & $\mp$ 0.001 & $\pm$ 0.005 \\
0.45 & 0.361 $\pm$ 0.002 $\pm$ 0.046 & $\pm$ 0.043 & $\mp$ 0.002 & $\pm$ 0.010 & $\pm$ 0.001 & $\mp$ 0.013 & $\pm$ 0.004 \\
0.55 & 0.241 $\pm$ 0.001 $\pm$ 0.043 & $\pm$ 0.040 & $\pm$ 0.002 & $\pm$ 0.006 & $\pm$ 0.006 & $\mp$ 0.014 & $\pm$ 0.002 \\ \hline\hline
\multicolumn{8}{|c|}{$110$ GeV $< \ptjet < 160$ GeV} \\ \hline
$r$ & $\rho \pm \ \rm (stat.) \pm \ \rm (syst.)$ &cluster e-scale & shower model &jet e-scale & resolution & correction & non-closure\\ \hline
0.05 & 6.292 $\pm$ 0.009 $\pm$ 0.160 & $\mp$ 0.095 & $\pm$ 0.067 & $\mp$ 0.056 & $\pm$ 0.018 & $\pm$ 0.067 & $\pm$ 0.063 \\
0.15 & 1.925 $\pm$ 0.005 $\pm$ 0.060 & $\mp$ 0.008 & $\mp$ 0.049 & $\pm$ 0.020 & $\pm$ 0.012 & $\mp$ 0.015 & $\pm$ 0.019 \\
0.25 & 0.830 $\pm$ 0.003 $\pm$ 0.043 & $\pm$ 0.020 & $\mp$ 0.023 & $\pm$ 0.016 & $\pm$ 0.001 & $\mp$ 0.024 & $\pm$ 0.008 \\
0.35 & 0.458 $\pm$ 0.002 $\pm$ 0.034 & $\pm$ 0.031 & $\mp$ 0.003 & $\pm$ 0.010 & $\pm$ 0.001 & $\mp$ 0.009 & $\pm$ 0.005 \\
0.45 & 0.292 $\pm$ 0.001 $\pm$ 0.035 & $\pm$ 0.033 & $\pm$ 0.003 & $\pm$ 0.008 & $\pm$ 0.002 & $\mp$ 0.009 & $\pm$ 0.003 \\
0.55 & 0.195 $\pm$ 0.001 $\pm$ 0.032 & $\pm$ 0.031 & $\pm$ 0.001 & $\pm$ 0.005 & $\pm$ 0.002 & $\mp$ 0.009 & $\pm$ 0.002 \\ \hline\hline
\multicolumn{8}{|c|}{$160$ GeV $< \ptjet < 210$ GeV} \\ \hline
$r$ & $\rho \pm \ \rm (stat.) \pm \ \rm (syst.)$ &cluster e-scale & shower model &jet e-scale & resolution & correction & non-closure\\ \hline

0.05 & 6.738 $\pm$ 0.012 $\pm$ 0.124 & $\mp$ 0.074 & $\pm$ 0.050 & $\mp$ 0.037 & $\pm$ 0.001 & $\pm$ 0.040 & $\pm$ 0.067 \\
0.15 & 1.722 $\pm$ 0.007 $\pm$ 0.055 & $\mp$ 0.002 & $\mp$ 0.046 & $\pm$ 0.018 & $\pm$ 0.002 & $\mp$ 0.019 & $\pm$ 0.017 \\
0.25 & 0.742 $\pm$ 0.005 $\pm$ 0.026 & $\pm$ 0.015 & $\mp$ 0.015 & $\pm$ 0.010 & $\pm$ 0.006 & $\mp$ 0.007 & $\pm$ 0.007 \\ 
0.35 & 0.394 $\pm$ 0.003 $\pm$ 0.025 & $\pm$ 0.024 & $\pm$ 0.003 & $\pm$ 0.004 & $\pm$ 0.003 & $\pm$ 0.001 & $\pm$ 0.004 \\
0.45 & 0.243 $\pm$ 0.002 $\pm$ 0.026 & $\pm$ 0.025 & $\pm$ 0.003 & $\pm$ 0.005 & $\pm$ 0.003 & $\mp$ 0.005 & $\pm$ 0.002 \\
0.55 & 0.155 $\pm$ 0.001 $\pm$ 0.024 & $\pm$ 0.023 & $\pm$ 0.002 & $\pm$ 0.002 & $\pm$ 0.002 & $\mp$ 0.007 & $\pm$ 0.002 \\ \hline\hline
\end{tabular}
\label{tab:table1}
\caption{\small
The measured differential jet shape, $\rho(r)$, as a function of $r$ in different $\ptjet$ regions,  
for jets with $|\rapjet| < 2.8$ and $30 \ {\rm GeV} < \ptjet < 210 \ {\rm GeV}$ (see Figs.~\ref{fig1} and ~\ref{fig1b}). 
The contributions from the different sources of systematic uncertainty are listed separately.}
\end{center}
\end{tiny}
\end{sidewaystable}


\begin{sidewaystable}
\begin{tiny}
\begin{center}
\begin{tabular}{|c|c||c|c|c|c|c|c|} \hline
\multicolumn{8}{|c|}{{\large{$\rho(r) \ (0 < |\rapjet|< 2.8)$}}} \\ \hline\hline\hline
\multicolumn{8}{|c|}{$210$ GeV $< \ptjet < 260$ GeV} \\ \hline
$r$ & $\rho \pm \ \rm (stat.) \pm \ \rm (syst.)$ &cluster e-scale & shower model &jet e-scale & resolution & correction & non-closure\\ \hline

0.05 & 7.004 $\pm$ 0.021 $\pm$ 0.146 & $\mp$ 0.061 & $\pm$ 0.084 & $\mp$ 0.037 & $\pm$ 0.055 & $\pm$ 0.035 & $\pm$ 0.070 \\
0.15 & 1.612 $\pm$ 0.012 $\pm$ 0.066 & $\mp$ 0.001 & $\mp$ 0.050 & $\pm$ 0.019 & $\pm$ 0.033 & $\mp$ 0.012 & $\pm$ 0.016 \\
0.25 & 0.672 $\pm$ 0.008 $\pm$ 0.035 & $\pm$ 0.012 & $\mp$ 0.027 & $\pm$ 0.011 & $\pm$ 0.013 & $\mp$ 0.005 & $\pm$ 0.007 \\
0.35 & 0.353 $\pm$ 0.005 $\pm$ 0.024 & $\pm$ 0.019 & $\mp$ 0.010 & $\pm$ 0.007 & $\pm$ 0.006 & $\mp$ 0.005 & $\pm$ 0.004 \\
0.45 & 0.212 $\pm$ 0.003 $\pm$ 0.024 & $\pm$ 0.020 & $\mp$ 0.007 & $\pm$ 0.005 & $\pm$ 0.004 & $\mp$ 0.008 & $\pm$ 0.002 \\
0.55 & 0.136 $\pm$ 0.001 $\pm$ 0.020 & $\pm$ 0.019 & $\mp$ 0.001 & $\pm$ 0.003 & $\pm$ 0.003 & $\mp$ 0.005 & $\pm$ 0.001 \\ \hline\hline
\multicolumn{8}{|c|}{$260$ GeV $< \ptjet < 310$ GeV} \\ \hline
$r$ & $\rho \pm \ \rm (stat.) \pm \ \rm (syst.)$ &cluster e-scale & shower model &jet e-scale & resolution & correction & non-closure\\ \hline
0.05 & 7.300 $\pm$ 0.036 $\pm$ 0.113 & $\mp$ 0.053 & $\pm$ 0.055 & $\mp$ 0.027 & $\pm$ 0.030 & $\mp$ 0.001 & $\pm$ 0.073 \\
0.15 & 1.463 $\pm$ 0.021 $\pm$ 0.038 & $\pm$ 0.001 & $\mp$ 0.030 & $\pm$ 0.016 & $\pm$ 0.009 & $\pm$ 0.004 & $\pm$ 0.015 \\
0.25 & 0.619 $\pm$ 0.014 $\pm$ 0.024 & $\pm$ 0.011 & $\mp$ 0.013 & $\pm$ 0.008 & $\pm$ 0.013 & $\pm$ 0.001 & $\pm$ 0.006 \\
0.35 & 0.315 $\pm$ 0.008 $\pm$ 0.019 & $\pm$ 0.016 & $\mp$ 0.010 & $\pm$ 0.002 & $\pm$ 0.003 & $\pm$ 0.003 & $\pm$ 0.003 \\
0.45 & 0.186 $\pm$ 0.004 $\pm$ 0.018 & $\pm$ 0.017 & $\mp$ 0.005 & $\pm$ 0.003 & $\pm$ 0.002 & $\pm$ 0.002 & $\pm$ 0.002 \\
0.55 & 0.115 $\pm$ 0.002 $\pm$ 0.016 & $\pm$ 0.015 & $\mp$ 0.001 & $\pm$ 0.002 & $\pm$ 0.004 & $\mp$ 0.004 & $\pm$ 0.001 \\ \hline\hline
\multicolumn{8}{|c|}{$310$ GeV $< \ptjet < 400$ GeV} \\ \hline
$r$ & $\rho \pm \ \rm (stat.) \pm \ \rm (syst.)$ &cluster e-scale & shower model &jet e-scale & resolution & correction & non-closure\\ \hline
0.05 & 7.495 $\pm$ 0.052 $\pm$ 0.128 & $\mp$ 0.043 & $\pm$ 0.059 & $\mp$ 0.034 & $\pm$ 0.033 & $\pm$ 0.056 & $\pm$ 0.075 \\
0.15 & 1.405 $\pm$ 0.031 $\pm$ 0.070 & $\pm$ 0.001 & $\mp$ 0.056 & $\pm$ 0.019 & $\pm$ 0.026 & $\mp$ 0.024 & $\pm$ 0.014 \\
0.25 & 0.536 $\pm$ 0.018 $\pm$ 0.023 & $\pm$ 0.008 & $\mp$ 0.008 & $\pm$ 0.006 & $\pm$ 0.001 & $\mp$ 0.019 & $\pm$ 0.005 \\
0.35 & 0.285 $\pm$ 0.011 $\pm$ 0.016 & $\pm$ 0.013 & $\pm$ 0.002 & $\pm$ 0.006 & $\pm$ 0.004 & $\mp$ 0.005 & $\pm$ 0.003 \\
0.45 & 0.173 $\pm$ 0.006 $\pm$ 0.016 & $\pm$ 0.014 & $\mp$ 0.001 & $\pm$ 0.003 & $\pm$ 0.005 & $\mp$ 0.004 & $\pm$ 0.002 \\
0.55 & 0.101 $\pm$ 0.003 $\pm$ 0.013 & $\pm$ 0.012 & $\pm$ 0.001 & $\pm$ 0.002 & $\pm$ 0.001 & $\mp$ 0.003 & $\pm$ 0.001 \\ \hline\hline
\multicolumn{8}{|c|}{$400$ GeV $< \ptjet < 500$ GeV} \\ \hline
$r$ & $\rho \pm \ \rm (stat.) \pm \ \rm (syst.)$ &cluster e-scale & shower model &jet e-scale & resolution & correction & non-closure\\ \hline
0.05 & 7.720 $\pm$ 0.114 $\pm$ 0.106 & $\mp$ 0.034 & $\pm$ 0.043 & $\mp$ 0.031 & $\pm$ 0.011 & $\pm$ 0.036 & $\pm$ 0.077 \\
0.15 & 1.339 $\pm$ 0.075 $\pm$ 0.054 & $\pm$ 0.001 & $\mp$ 0.047 & $\pm$ 0.020 & $\pm$ 0.010 & $\mp$ 0.003 & $\pm$ 0.013 \\
0.25 & 0.489 $\pm$ 0.039 $\pm$ 0.023 & $\pm$ 0.006 & $\pm$ 0.001 & $\pm$ 0.008 & $\pm$ 0.005 & $\mp$ 0.020 & $\pm$ 0.005 \\
0.35 & 0.226 $\pm$ 0.019 $\pm$ 0.012 & $\pm$ 0.009 & $\mp$ 0.003 & $\pm$ 0.003 & $\pm$ 0.003 & $\mp$ 0.005 & $\pm$ 0.002 \\
0.45 & 0.128 $\pm$ 0.009 $\pm$ 0.011 & $\pm$ 0.010 & $\pm$ 0.001 & $\pm$ 0.002 & $\pm$ 0.002 & $\mp$ 0.003 & $\pm$ 0.001 \\
0.55 & 0.086 $\pm$ 0.006 $\pm$ 0.011 & $\pm$ 0.010 & $\pm$ 0.001 & $\pm$ 0.001 & $\pm$ 0.001 & $\mp$ 0.003 & $\pm$ 0.001 \\ \hline\hline 
\multicolumn{8}{|c|}{$500$ GeV $< \ptjet < 600$ GeV} \\ \hline
$r$ & $\rho \pm \ \rm (stat.) \pm \ \rm (syst.)$ &cluster e-scale & shower model &jet e-scale & resolution & correction & non-closure\\ \hline
0.05 & 7.638 $\pm$ 0.261 $\pm$ 0.093 & $\mp$ 0.026 & $\pm$ 0.001 & $\mp$ 0.022 & $\pm$ 0.009 & $\pm$ 0.040 & $\pm$ 0.076 \\
0.15 & 1.400 $\pm$ 0.168 $\pm$ 0.037 & $\pm$ 0.001 & $\mp$ 0.011 & $\pm$ 0.013 & $\pm$ 0.003 & $\mp$ 0.030 & $\pm$ 0.014 \\
0.25 & 0.475 $\pm$ 0.074 $\pm$ 0.017 & $\pm$ 0.006 & $\pm$ 0.008 & $\pm$ 0.009 & $\pm$ 0.010 & $\mp$ 0.003 & $\pm$ 0.005 \\
0.35 & 0.257 $\pm$ 0.054 $\pm$ 0.012 & $\pm$ 0.010 & $\mp$ 0.004 & $\pm$ 0.004 & $\pm$ 0.001 & $\mp$ 0.002 & $\pm$ 0.003 \\
0.45 & 0.153 $\pm$ 0.037 $\pm$ 0.012 & $\pm$ 0.011 & $\pm$ 0.002 & $\pm$ 0.004 & $\pm$ 0.001 & $\mp$ 0.002 & $\pm$ 0.002 \\
0.55 & 0.078 $\pm$ 0.015 $\pm$ 0.010 & $\pm$ 0.009 & $\pm$ 0.004 & $\pm$ 0.002 & $\pm$ 0.001 & $\mp$ 0.003 & $\pm$ 0.001 \\ \hline\hline
\end{tabular}
\label{tab:table2}
\caption{\small
The measured differential jet shape, $\rho(r)$, as a function of $r$ in different $\ptjet$ regions,  
for jets with $|\rapjet| < 2.8$ and $210 \ {\rm GeV} < \ptjet < 600 \ {\rm GeV}$ (see Fig.~\ref{fig2}). 
The contributions from the different sources of systematic uncertainty are listed separately.}
\end{center}
\end{tiny}
\end{sidewaystable}


\begin{sidewaystable}
\begin{tiny}
\begin{center}
\begin{tabular}{|c|c||c|c|c|c|c|c|} \hline
\multicolumn{8}{|c|}{{\large{$1-\Psi(r=0.3)$}}} \\ \hline\hline\hline
\multicolumn{8}{|c|}{$(0 < |\rapjet|<2.8)$} \\ \hline
$\ptjet$ (GeV)& $1-\Psi(r=0.3) \pm \ \rm (stat.) \pm \ \rm (syst.)$ &cluster e-scale & shower model &jet e-scale & resolution & correction & non-closure\\ \hline
30 - 40 & 0.2193 $\pm$ 0.0006 $\pm$ 0.0325 & $\pm$ 0.0212 & $\pm$ 0.0001 & $\pm$ 0.0105 & $\pm$ 0.0057 & $\pm$ 0.0216 &-\\
40 - 60 & 0.1733 $\pm$ 0.0008 $\pm$ 0.0221 & $\pm$ 0.0177 & $\pm$ 0.0006 & $\pm$ 0.0070 & $\pm$ 0.0041 & $\pm$ 0.0104 &-\\
60 - 80 & 0.1347 $\pm$ 0.0004 $\pm$ 0.0157 & $\pm$ 0.0138 & $\pm$ 0.0010 & $\pm$ 0.0035 & $\pm$ 0.0017 & $\pm$ 0.0064 &-\\
80 - 110 & 0.1146 $\pm$ 0.0003 $\pm$ 0.0117 & $\pm$ 0.0109 & $\pm$ 0.0005 & $\pm$ 0.0025 & $\pm$ 0.0007 & $\pm$ 0.0033 &-\\
110 - 160 & 0.0942 $\pm$ 0.0003 $\pm$ 0.0092 & $\pm$ 0.0084 & $\pm$ 0.0001 & $\pm$ 0.0021 & $\pm$ 0.0005 & $\pm$ 0.0030 &-\\
160 - 210 & 0.0789 $\pm$ 0.0004 $\pm$ 0.0067 & $\pm$ 0.0063 & $\pm$ 0.0007 & $\pm$ 0.0010 & $\pm$ 0.0008 & $\pm$ 0.0015 &-\\
210 - 260 & 0.0698 $\pm$ 0.0006 $\pm$ 0.0059 & $\pm$ 0.0051 & $\pm$ 0.0015 & $\pm$ 0.0013 & $\pm$ 0.0011 & $\pm$ 0.0020 &-\\
260 - 310 & 0.0615 $\pm$ 0.0010 $\pm$ 0.0046 & $\pm$ 0.0042 & $\pm$ 0.0014 & $\pm$ 0.0006 & $\pm$ 0.0008 & $\pm$ 0.0003 &-\\
310 - 400 & 0.0556 $\pm$ 0.0015 $\pm$ 0.0041 & $\pm$ 0.0035 & $\pm$ 0.0001 & $\pm$ 0.0010 & $\pm$ 0.0007 & $\pm$ 0.0016 &-\\
400 - 500 & 0.0442 $\pm$ 0.0024 $\pm$ 0.0033 & $\pm$ 0.0028 & $\pm$ 0.0001 & $\pm$ 0.0006 & $\pm$ 0.0006 & $\pm$ 0.0016 &-\\
500 - 600 & 0.0479 $\pm$ 0.0070 $\pm$ 0.0026 & $\pm$ 0.0022 & $\pm$ 0.0002 & $\pm$ 0.0008 & $\pm$ 0.0001 & $\pm$ 0.0012 &-\\ \hline\hline
\end{tabular}
\label{tab:table3}
\caption{\small
The measured integrated  jet shape, $1-\Psi(r=0.3)$, as a function of $\ptjet$, for jets
with $|\rapjet| < 2.8$ and $30 \ {\rm GeV} < \ptjet < 600 \ {\rm GeV}$ (see Fig.~\ref{fig3}). 
The contributions from the different sources of systematic uncertainty are listed separately.}
\end{center}
\end{tiny}
\end{sidewaystable}


\begin{sidewaystable}
\begin{tiny}
\begin{center}
\begin{tabular}{|c|c||c|c|c|c|c|c|} \hline
\multicolumn{8}{|c|}{{\large{$1-\Psi(r=0.3)$}}} \\ \hline\hline\hline
\multicolumn{8}{|c|}{$(0 < |\rapjet|<0.3)$} \\ \hline
$\ptjet$ (GeV)& $1-\Psi(r=0.3) \pm \ \rm (stat.) \pm \ \rm (syst.)$ &cluster e-scale & shower model &jet e-scale & resolution & correction & non-closure\\ \hline
30 - 40 & 0.2175 $\pm$ 0.0016 $\pm$ 0.0309 & $\pm$ 0.0149 & $\pm$ 0.0050 & $\pm$ 0.0112 & $\pm$ 0.0057 & $\pm$ 0.0234 &-\\
40 - 60 & 0.1751 $\pm$ 0.0023 $\pm$ 0.0249 & $\pm$ 0.0133 & $\pm$ 0.0002 & $\pm$ 0.0074 & $\pm$ 0.0041 & $\pm$ 0.0192 &-\\
60 - 80 & 0.1395 $\pm$ 0.0011 $\pm$ 0.0147 & $\pm$ 0.0105 & $\pm$ 0.0070 & $\pm$ 0.0039 & $\pm$ 0.0017 & $\pm$ 0.0062 &-\\
80 - 110 & 0.1203 $\pm$ 0.0009 $\pm$ 0.0110 & $\pm$ 0.0086 & $\pm$ 0.0035 & $\pm$ 0.0022 & $\pm$ 0.0007 & $\pm$ 0.0055 &-\\
110 - 160 & 0.0990 $\pm$ 0.0007 $\pm$ 0.0087 & $\pm$ 0.0067 & $\pm$ 0.0025 & $\pm$ 0.0017 & $\pm$ 0.0005 & $\pm$ 0.0047 &-\\
160 - 210 & 0.0831 $\pm$ 0.0010 $\pm$ 0.0074 & $\pm$ 0.0051 & $\pm$ 0.0004 & $\pm$ 0.0008 & $\pm$ 0.0008 & $\pm$ 0.0053 &-\\
210 - 260 & 0.0758 $\pm$ 0.0015 $\pm$ 0.0047 & $\pm$ 0.0042 & $\pm$ 0.0017 & $\pm$ 0.0008 & $\pm$ 0.0011 & $\pm$ 0.0006 &-\\
260 - 310 & 0.0639 $\pm$ 0.0024 $\pm$ 0.0068 & $\pm$ 0.0035 & $\pm$ 0.0032 & $\pm$ 0.0003 & $\pm$ 0.0008 & $\pm$ 0.0049 &-\\
310 - 400 & 0.0578 $\pm$ 0.0031 $\pm$ 0.0034 & $\pm$ 0.0030 & $\pm$ 0.0002 & $\pm$ 0.0013 & $\pm$ 0.0007 & $\pm$ 0.0007 &-\\
400 - 500 & 0.0486 $\pm$ 0.0044 $\pm$ 0.0037 & $\pm$ 0.0024 & $\pm$ 0.0022 & $\pm$ 0.0006 & $\pm$ 0.0006 & $\pm$ 0.0017 &-\\ \hline\hline
\multicolumn{8}{|c|}{$(0.3 < |\rapjet|<0.8)$} \\ \hline
$\ptjet$ (GeV)& $1-\Psi(r=0.3) \pm \ \rm (stat.) \pm \ \rm (syst.)$ &cluster e-scale & shower model &jet e-scale & resolution & correction & non-closure\\ \hline 
30 - 40 & 0.2219 $\pm$ 0.0012 $\pm$ 0.0390 & $\pm$ 0.0173 & $\pm$ 0.0036 & $\pm$ 0.0109 & $\pm$ 0.0057 & $\pm$ 0.0326 &-\\
40 - 60 & 0.1779 $\pm$ 0.0017 $\pm$ 0.0233 & $\pm$ 0.0145 & $\pm$ 0.0051 & $\pm$ 0.0059 & $\pm$ 0.0041 & $\pm$ 0.0160 &-\\
60 - 80 & 0.1378 $\pm$ 0.0008 $\pm$ 0.0159 & $\pm$ 0.0117 & $\pm$ 0.0021 & $\pm$ 0.0041 & $\pm$ 0.0017 & $\pm$ 0.0097 &-\\
80 - 110 & 0.1179 $\pm$ 0.0007 $\pm$ 0.0116 & $\pm$ 0.0093 & $\pm$ 0.0002 & $\pm$ 0.0025 & $\pm$ 0.0007 & $\pm$ 0.0063 &-\\
110 - 160 & 0.0963 $\pm$ 0.0006 $\pm$ 0.0094 & $\pm$ 0.0073 & $\pm$ 0.0006 & $\pm$ 0.0018 & $\pm$ 0.0005 & $\pm$ 0.0056 &-\\
160 - 210 & 0.0847 $\pm$ 0.0007 $\pm$ 0.0061 & $\pm$ 0.0055 & $\pm$ 0.0017 & $\pm$ 0.0017 & $\pm$ 0.0008 & $\pm$ 0.0011 &-\\ 
210 - 260 & 0.0718 $\pm$ 0.0012 $\pm$ 0.0067 & $\pm$ 0.0045 & $\pm$ 0.0023 & $\pm$ 0.0016 & $\pm$ 0.0011 & $\pm$ 0.0039 &-\\
260 - 310 & 0.0631 $\pm$ 0.0019 $\pm$ 0.0042 & $\pm$ 0.0038 & $\pm$ 0.0009 & $\pm$ 0.0008 & $\pm$ 0.0008 & $\pm$ 0.0010 &-\\
310 - 400 & 0.0623 $\pm$ 0.0030 $\pm$ 0.0042 & $\pm$ 0.0031 & $\pm$ 0.0016 & $\pm$ 0.0011 & $\pm$ 0.0007 & $\pm$ 0.0019 &-\\
400 - 500 & 0.0384 $\pm$ 0.0033 $\pm$ 0.0042 & $\pm$ 0.0025 & $\pm$ 0.0005 & $\pm$ 0.0007 & $\pm$ 0.0006 & $\pm$ 0.0031 &-\\ \hline\hline
\multicolumn{8}{|c|}{$(0.8 < |\rapjet|<1.2)$} \\ \hline
$\ptjet$ (GeV)& $1-\Psi(r=0.3) \pm \ \rm (stat.) \pm \ \rm (syst.)$ &cluster e-scale & shower model &jet e-scale & resolution & correction & non-closure\\ \hline
30 - 40 & 0.2191  $\pm$ 0.0014  $\pm$ 0.0314 & $\pm$ 0.0233 & $\pm$ 0.0030 & $\pm$ 0.0102 & $\pm$ 0.0057 & $\pm$ 0.0172 &-\\
40 - 60 & 0.1736  $\pm$ 0.0020  $\pm$ 0.0247 & $\pm$ 0.0192 & $\pm$ 0.0030 & $\pm$ 0.0090 & $\pm$ 0.0041 & $\pm$ 0.0116 &-\\
60 - 80 & 0.1347  $\pm$ 0.0009  $\pm$ 0.0173 & $\pm$ 0.0151 & $\pm$ 0.0013 & $\pm$ 0.0052 & $\pm$ 0.0017 & $\pm$ 0.0063 &-\\
80 - 110 & 0.1161  $\pm$ 0.0008  $\pm$ 0.0133 & $\pm$ 0.0118 & $\pm$ 0.0001 & $\pm$ 0.0034 & $\pm$ 0.0007 & $\pm$ 0.0051 &-\\
110 - 160 & 0.0975  $\pm$ 0.0007  $\pm$ 0.0105 & $\pm$ 0.0092 & $\pm$ 0.0001 & $\pm$ 0.0024 & $\pm$ 0.0005 & $\pm$ 0.0043 &-\\ 
160 - 210 & 0.0817  $\pm$ 0.0009  $\pm$ 0.0071 & $\pm$ 0.0069 & $\pm$ 0.0007 & $\pm$ 0.0014 & $\pm$ 0.0008 & $\pm$ 0.0010 &-\\
210 - 260 & 0.0721  $\pm$ 0.0016  $\pm$ 0.0073 & $\pm$ 0.0054 & $\pm$ 0.0010 & $\pm$ 0.0015 & $\pm$ 0.0011 & $\pm$ 0.0044 &-\\
260 - 310 & 0.0639  $\pm$ 0.0022  $\pm$ 0.0051 & $\pm$ 0.0046 & $\pm$ 0.0010 & $\pm$ 0.0016 & $\pm$ 0.0008 & $\pm$ 0.0002 &-\\
310 - 400 & 0.0529  $\pm$ 0.0031  $\pm$ 0.0058 & $\pm$ 0.0038 & $\pm$ 0.0001 & $\pm$ 0.0009 & $\pm$ 0.0007 & $\pm$ 0.0042 &-\\
400 - 500 & 0.0593  $\pm$ 0.0079  $\pm$ 0.0037 & $\pm$ 0.0030 & $\pm$ 0.0014 & $\pm$ 0.0011 & $\pm$ 0.0006 & $\pm$ 0.0012 &-\\ \hline\hline
\end{tabular}
\label{tab:table4}
\caption{\small
The measured integrated  jet shape, $1-\Psi(r=0.3)$, as a function of $\ptjet$, for jets
with $30 \ {\rm GeV} < \ptjet < 500 \ {\rm GeV}$ in different jet rapidity regions 
(see Fig.~\ref{fig6}). The contributions from the different sources of systematic 
uncertainty are listed separately.
}
\end{center}
\end{tiny}
\end{sidewaystable}


\begin{sidewaystable}
\begin{tiny}
\begin{center}
\begin{tabular}{|c|c||c|c|c|c|c|c|} \hline
\multicolumn{8}{|c|}{{\large{$1-\Psi(r=0.3)$}}} \\ \hline\hline\hline
\multicolumn{8}{|c|}{$(1.2 < |\rapjet|<2.1)$} \\ \hline
$\ptjet$ (GeV)& $1-\Psi(r=0.3) \pm \ \rm (stat.) \pm \ \rm (syst.)$ &cluster e-scale & shower model &jet e-scale & resolution & correction & non-closure\\ \hline
30 - 40 & 0.2177 $\pm$ 0.0010 $\pm$ 0.0325 & $\pm$ 0.0263 & $\pm$ 0.0017 & $\pm$ 0.0114 & $\pm$ 0.0057 & $\pm$ 0.0140 &-\\
40 - 60 & 0.1731 $\pm$ 0.0014 $\pm$ 0.0244 & $\pm$ 0.0217 & $\pm$ 0.0014 & $\pm$ 0.0066 & $\pm$ 0.0041 & $\pm$ 0.0077 &-\\
60 - 80 & 0.1331 $\pm$ 0.0007 $\pm$ 0.0178 & $\pm$ 0.0168 & $\pm$ 0.0001 & $\pm$ 0.0035 & $\pm$ 0.0017 & $\pm$ 0.0045 &-\\
80 - 110 & 0.1130 $\pm$ 0.0006 $\pm$ 0.0140 & $\pm$ 0.0133 & $\pm$ 0.0029 & $\pm$ 0.0029 & $\pm$ 0.0007 & $\pm$ 0.0012 &-\\
110 - 160 & 0.0904 $\pm$ 0.0005 $\pm$ 0.0109 & $\pm$ 0.0103 & $\pm$ 0.0010 & $\pm$ 0.0019 & $\pm$ 0.0005 & $\pm$ 0.0029 &-\\
160 - 210 & 0.0735 $\pm$ 0.0007 $\pm$ 0.0082 & $\pm$ 0.0077 & $\pm$ 0.0011 & $\pm$ 0.0015 & $\pm$ 0.0008 & $\pm$ 0.0019 &-\\
210 - 260 & 0.0646 $\pm$ 0.0011 $\pm$ 0.0066 & $\pm$ 0.0061 & $\pm$ 0.0007 & $\pm$ 0.0014 & $\pm$ 0.0011 & $\pm$ 0.0014 &-\\
260 - 310 & 0.0573 $\pm$ 0.0021 $\pm$ 0.0053 & $\pm$ 0.0051 & $\pm$ 0.0007 & $\pm$ 0.0011 & $\pm$ 0.0008 & $\pm$ 0.0002 &-\\
310 - 400 & 0.0495 $\pm$ 0.0026 $\pm$ 0.0045 & $\pm$ 0.0043 & $\pm$ 0.0005 & $\pm$ 0.0008 & $\pm$ 0.0007 & $\pm$ 0.0009 &-\\ 
400 - 500 & 0.0335 $\pm$ 0.0033 $\pm$ 0.0037 & $\pm$ 0.0035 & $\pm$ 0.0006 & $\pm$ 0.0007 & $\pm$ 0.0006 & $\pm$ 0.0006 &-\\ \hline\hline
\multicolumn{8}{|c|}{$(2.1 < |\rapjet|<2.8)$} \\ \hline
$\ptjet$ (GeV)& $1-\Psi(r=0.3) \pm \ \rm (stat.) \pm \ \rm (syst.)$ &cluster e-scale & shower model &jet e-scale & resolution & correction & non-closure\\ \hline
30 - 40 & 0.2110 $\pm$ 0.0014 $\pm$ 0.0256 & $\pm$ 0.0209 & $\pm$ 0.0094 & $\pm$ 0.0098 & $\pm$ 0.0057 & $\pm$ 0.0003 &-\\
40 - 60 & 0.1664 $\pm$ 0.0021 $\pm$ 0.0193 & $\pm$ 0.0169 & $\pm$ 0.0048 & $\pm$ 0.0066 & $\pm$ 0.0042 & $\pm$ 0.0023 &-\\
60 - 80 & 0.1274 $\pm$ 0.0011 $\pm$ 0.0153 & $\pm$ 0.0126 & $\pm$ 0.0062 & $\pm$ 0.0057 & $\pm$ 0.0017 & $\pm$ 0.0012 &-\\
80 - 110 & 0.1048 $\pm$ 0.0009 $\pm$ 0.0110 & $\pm$ 0.0099 & $\pm$ 0.0031 & $\pm$ 0.0033 & $\pm$ 0.0007 & $\pm$ 0.0004 &-\\
110 - 160 & 0.0830 $\pm$ 0.0008 $\pm$ 0.0090 & $\pm$ 0.0076 & $\pm$ 0.0026 & $\pm$ 0.0034 & $\pm$ 0.0005 & $\pm$ 0.0019 &-\\
160 - 210 & 0.0626 $\pm$ 0.0010 $\pm$ 0.0074 & $\pm$ 0.0058 & $\pm$ 0.0030 & $\pm$ 0.0026 & $\pm$ 0.0008 & $\pm$ 0.0020 &-\\
210 - 260 & 0.0607 $\pm$ 0.0023 $\pm$ 0.0066 & $\pm$ 0.0048 & $\pm$ 0.0018 & $\pm$ 0.0027 & $\pm$ 0.0011 & $\pm$ 0.0029 &-\\
260 - 310 & 0.0538 $\pm$ 0.0040 $\pm$ 0.0047 & $\pm$ 0.0040 & $\pm$ 0.0022 & $\pm$ 0.0007 & $\pm$ 0.0009 & $\pm$ 0.0006 &-\\ \hline\hline
\end{tabular}
\label{tab:table5}
\caption{\small
The measured integrated  jet shape, $1-\Psi(r=0.3)$, as a function of $\ptjet$, for jets
with $30 \ {\rm GeV} < \ptjet < 500 \ {\rm GeV}$ in different jet rapidity regions 
(see Fig.~\ref{fig6}). The contributions from the different sources of systematic 
uncertainty are listed separately.
}
\end{center}
\end{tiny}
\end{sidewaystable}
%


 \begin{sidewaystable}
 \begin{footnotesize}
 \begin{center}
 \begin{tabular}{|c||c|c|c|c|c|} \hline
 \multicolumn{6}{|c|}{{\large{$\chi^2/d.o.f$}}} \\ \hline\hline\hline
   & $0 < |\rapjet| < 0.3$ &  $0.3 < |\rapjet| < 0.8$ & $0.8 < |\rapjet| < 1.2$ & $1.2 < |\rapjet| < 2.1$ & $2.1 < |\rapjet| < 2.8$ \\ \hline\hline
degrees of freedom (d.o.f)               &   10   &   10    &   10    &  10     & 8    \\ \hline 
PYTHIA-Perugia2010  & 0.6  &  1.8  &  2.4  &  1.4  &  1.4 \\
HERWIG++            & 2.2  &  2.3  &  3.1  &  1.8  &  4.0 \\
PYTHIA-MC09         & 1.0  &  2.5  &  2.4  &  1.5  &  3.2 \\
PYTHIA-DW           & 2.4  &  3.4  &  6.9 &  4.0  &   5.2 \\
ALPGEN              & 3.8  &  9.8  &   7.4 &   6.7 &   6.0  \\
PYTHIA-Perugia2010 (no UE) &  4.2 &   9.7 &   4.9 &   8.6 &   4.8 \\ \hline\hline
\end{tabular}
\label{tab:chi2}
\caption{Results of $\chi^2$ tests to the data in Fig.~\ref{fig6} with 
respect to the different MC predictions. As discussed in the text, the 
different sources of systematic uncertainty are considered 
independent and fully correlated across $\ptjet$ bins.}
\end{center}
\end{footnotesize}
\end{sidewaystable}


\clearpage
\begin{flushleft}
{\Large The ATLAS Collaboration}

\bigskip

G.~Aad$^{\rm 48}$,
B.~Abbott$^{\rm 111}$,
J.~Abdallah$^{\rm 11}$,
A.A.~Abdelalim$^{\rm 49}$,
A.~Abdesselam$^{\rm 118}$,
O.~Abdinov$^{\rm 10}$,
B.~Abi$^{\rm 112}$,
M.~Abolins$^{\rm 88}$,
H.~Abramowicz$^{\rm 153}$,
H.~Abreu$^{\rm 115}$
% ...
\bigskip

% ...
$^{10}$ Institute of Physics, Azerbaijan Academy of Sciences, H. Javid Avenue 33, AZ 143 Baku, Azerbaijan\\
$^{11}$ Institut de F\'isica d'Altes Energies, IFAE, Edifici Cn, Universitat Aut\`onoma  de Barcelona,  ES - 08193 Bellaterra (Barcelona), Spain\\
% ...
$^{48}$ Albert-Ludwigs-Universit\"{a}t, Fakult\"{a}t f\"{u}r Mathematik und Physik, Hermann-Herder Str. 3, D - 79104 Freiburg i.Br., Germany\\
$^{49}$ Universit\'e de Gen\`eve, Section de Physique, 24 rue Ernest Ansermet, CH - 1211 Geneve 4, Switzerland\\
% ...
$^{88}$ Michigan State University, Department of Physics and Astronomy, High Energy Physics Group, East Lansing, MI 48824-2320, United States of America\\
% ...
$^{111}$ University of Oklahoma, Homer L. Dodge Department of Physics and Astronomy, 440 West Brooks, Room 100, Norman, OK 73019-0225, United States of America\\
$^{112}$ Oklahoma State University, Department of Physics, 145 Physical Sciences Building, Stillwater, OK 74078-3072, United States of America\\
% ...
$^{115}$ LAL, Univ. Paris-Sud, IN2P3/CNRS, Orsay, France\\
% ...
$^{118}$ Oxford University, Department of Physics, Denys Wilkinson Building, Keble Road, Oxford OX1 3RH, United Kingdom\\
% ...
$^{153}$ Tel Aviv University, Raymond and Beverly Sackler School of Physics and Astronomy, Ramat Aviv, IL - Tel Aviv 69978, Israel\\
% ...
\end{flushleft}


\begin{thebibliography}{99.}
\bibitem{jetshape} S. D. Ellis, Z. Kunszt and D. E. Soper, Phys. Rev. Lett. {\bf 69} 3615 (1992).
\bibitem{pqcd}D. J. Gross and F. Wilczek, Phys. Rev. D {\bf 8} 3633 (1973).
\bibitem{bjet}Inclusive jet shape studies  
have a very limited sensitivity to the presence of 
a small contribution from heavy-flavor quarks in the final state.
\bibitem{jet_pro} The CDF Collaboration, A.~Abulencia {\it et al.}, Phys. Rev. D {\bf 75} 092006 (2007).\\
                  The D0  Collaboration, V.~M. Abazov {\it et al.}, Phys. Rev. Lett. {\bf 101} 062001 (2008).\\
                  The CDF Collaboration, T.~Aaltonen {\it et al.},  Phys.  Rev. D {\bf 78}      052006 (2008).
\bibitem{jet_pro_atlas} The ATLAS Collaboration, G. Aad {\it et al.}, CERN-PH-EP-2010-034; arXiv:1009.5908 (2010); accepted for publication in Eur. Phys. J. C, and references therein.
\bibitem{boosted}  J. M. Butterworth, A. R. Davison, M. Rubin and G. P. Salam, Phys. Rev. Lett. {\bf 100} 242001 (2008).\\
                   D. Kaplan {\it et al.}, Phys. Rev. Lett. {\bf 101} 142001 (2008).\\
                   G. Salam, Eur. Phys. J. C {\bf 67} 637 (2010). 
\bibitem{ppbar} The CDF Collaboration,  D. Acosta {\it et al.}, Phys. Rev. D {\bf 71} 112002 (2005).\\
             The CDF Collaboration, F. Abe {\it et al.}, Phys. Rev. Lett. {\bf 70} 713 (1993). \\
             The D0 Collaboration, S. Abachi {\it et al.}, Phys. Lett. B {\bf 357} 500 (1995).
\bibitem{ep} The ZEUS Collaboration, S. Chekanov {\it et al.}, Nucl. Phys. B {\bf 700} 3 (2004).\\
             The ZEUS Collaboration, J.Breitweg {\it et al.}, Eur. Phys. J. C {\bf 8} 3 367 (1999).\\
             The H1 Collaboration, C. Adloff {\it et al.}, Nucl. Phys. B {\bf 545} 3 (1999).\\ 
             The ZEUS Collaboration, J.Breitweg {\it et al.}, Eur. Phys. J. C {\bf 2} 1 61 (1998). 
\bibitem{ee} The OPAL Collaboration, R. Akers {\it et al.}, Z. Phys. C {\bf 63} 197 (1994).\\
             The OPAL Collaboration, K. Ackerstaff {\it et al.}, Eur. Phys. J. C {\bf 1} 479 (1998).
\bibitem{atlas} The ATLAS Collaboration, G. Aad {\it et al.}, JINST {\bf 3} S08003 (2008).


\bibitem{coord} 
The ATLAS reference system is a Cartesian 
right-handed coordinate system, with the nominal collision point at the origin. The anti-clockwise beam 
direction defines the positive $z$-axis, while the positive $x$-axis is defined as pointing from the collision 
point to the centre of the LHC ring and the positive $y$-axis points upwards.  The azimuthal angle $\phi$ is 
measured around the beam axis, and the polar angle $\theta$ is measured with respect to the $z$-axis.
The pseudorapidity is defined as $\eta = - {\rm ln}({\rm tan}(\theta/2))$. 
The rapidity is defined as $y = 0.5 \times {\rm ln}[(E+p_z)/(E-p_z)]$, where $E$ denotes the energy and $p_z$ is the component of the momentum along the beam direction.

\bibitem{mbts}  The ATLAS Collaboration,  G. Aad {\it et al.}, Phys. Lett. B {\bf 688} 21 (2010).
\bibitem{PYTHIA} T. Sj\"ostrand  {\it et al.}, JHEP {\bf 05} 026 (2006).
\bibitem{HERWIG++} M. Bahr {\it et al.}, HERWIG++ Physics and Manual, Eur. Phys. J. C~{\bf 58} 639 (2008).
\bibitem{string} B. Andersson {\it et al.}, Phys. Rep. {\bf 97} 31 (1983).  
\bibitem{cluster} B.R. Webber, Nucl. Phys. B {\bf 238} 492 (1984).
\bibitem{MC09} ATLAS Collaboration, ATLAS MC tunes for MC09, ATL-PHYS-PUB-2010-002 (2010).
\bibitem{DW} The CDF Collaboration, T. Aaltonen {\it et al.}, Phys. Rev. D~{\bf 82} 034001 (2010).
\bibitem{Perugia2010} P. Z. Skands, CERN-PH-TH-2010-113, arXiv:hep-ph/1005.3457 (2010).
\bibitem{ALPGEN} M.L. Mangano {\it et al.}, JHEP {\bf 01} 0307 (2003).
\bibitem{herwig}  G. Corcella {\it et al.}, JHEP {\bf 0101} 010 (2001).  
\bibitem{Jimmy}J. Butterworth, J. Forshaw and M.Seymour, Z. Phys. C~{\bf 72} 637 (1996).
\bibitem{mrst2007lo} A.~D. Martin, W.~J. Stirling, R.~S. Thorne and G.~Watt, 
                     Eur. Phys. J. C {\bf 63} 189 (2009).\\
 A.~Sherstnev and R.~S. Thorne, Eur. Phys. J. C {\bf 55} 553 (2008).
\bibitem{cteq}  J. Pumplin {\it et al.}, JHEP {\bf 0207} 012 (2002).
\bibitem{cteq6}  D. Stump {\it et al.}, JHEP { \bf 0310 } 046 (2003).
\bibitem{atlas_sim} The ATLAS Collaboration,  G. Aad {\it et al.}, Eur. Phys. J. C~{\bf 70} 823 (2010).
\bibitem{geant}S. Agostinelli {\it et al.}, Nucl.  Instrum.  and Meth. {\bf A506} 250 (2003).
\bibitem{qgs} G. Folger and J.P. Wellisch, arXiv:nucl-th/0306007 (2003). 
\bibitem{bertini} H. Bertini, Phys. Rev. {\bf 188} 1711 (1969).  
\bibitem{sim_barrel} E.~Abat {\it et al.}, Tech. Rep. ATL-CAL-PUB-2010-001, CERN, Geneva, (2010).\\
P.~Adragna {\it et al.}, CERN-PH-EP-2009-019; ATL-TILECAL-PUB-2009-009 (2009).\\
E.~Abat {\it et al.}, Nucl. Instrum. and Meth. {\bf A607} 372 (2009). \\
E.~Abat {\it et al.}, Nucl. Instrum. and Meth. {\bf A615} 158 (2010). \\
E.~Abat {\it et al.}, Nucl. Instrum. and Meth.  {\bf A621} 134 (2010). 

\bibitem{sim_endcap} J.~Pinfold {\it et al.}, Nucl. Instrum. Meth. {\bf A593} 324 (2008).\\
D.~M. Gingrich et al., J. Inst. {\bf 2} no.~05 P05005 (2007).

\bibitem{akt}M. Cacciari, G. P. Salam and G. Soyez, JHEP 0804 063 (2008).
\bibitem{particle}  The final state in the MC generators is defined 
using all particles (including muons and neutrinos) with lifetime above $10^{-11} {\rm s}$. 
\bibitem{emscale}The electromagnetic scale is the appropriate scale for the reconstruction of the energy deposited by electrons or photons in the calorimeter.
\bibitem{ftfp}  B. Andersson, G. Gustafson and B. Nilsson-Almqvist, Nucl. Phys. B~{\bf 281} 289 (1987).
\bibitem{tab} A complete set of tables for differential and integrated measurements
as a function of $\ptjet$ and $|\rapjet|$ 
are available at the Durham HepData repository 
(http://hepdata.cedar.ac.uk). 


\end{thebibliography}
\end{document}